\documentclass{article}

\usepackage{PRIMEarxiv}

\usepackage{acronym}
\usepackage{algpseudocode}
\usepackage{algorithm}
\usepackage{amsmath,amsfonts}
\usepackage{array}
\usepackage{booktabs}

\usepackage{doi}
\usepackage[T1]{fontenc}
\usepackage[acronym]{glossaries}
\usepackage{graphicx}
\usepackage[utf8]{inputenc}
\usepackage{lipsum}
\usepackage{makecell}
\usepackage{mathtools}
\mathtoolsset{showonlyrefs}
\usepackage{microtype}
\usepackage{multicol}
\usepackage{multirow}
\usepackage[numbers]{natbib}
\usepackage{nicefrac}
\usepackage{stfloats}
\usepackage{enumitem}
\usepackage{subcaption}
\usepackage{textcomp}
\usepackage{threeparttable}
\usepackage{url}
\usepackage{verbatim}
\usepackage{stackengine}
\usepackage{xcolor}
\usepackage{todonotes}
\usepackage{CJKutf8}
\usepackage{siunitx}
\usepackage{adjustbox}
\usepackage{rotating}

\title{Neural Machine Unranking
\thanks{\textit{\underline{Corresponding author}}: 
\textbf{Georgina Cosma}} 
}

\author{
  Jingrui Hou \\
  School of Information Management,
  Wuhan University,
  Wuhan, Hubei, China \\
  Department of Computer Science,
  Loughborough University,
  Loughborough, Leicestershire, UK \\
  \texttt{houjingrui@whu.edu.cn} \\
  \AND
  Axel Finke \\
  Department of Mathematical Sciences \\
  Loughborough University \\
  Loughborough, Leicestershire, UK \\
  \texttt{a.finke@lboro.ac.uk} \\
  \And
  Georgina Cosma \\
  Department of Computer Science \\
  Loughborough University \\
  Loughborough, Leicestershire, UK \\
  \texttt{g.cosma@lboro.ac.uk} \\
}

\newcommand{\singleDocument}{d}
\newcommand{\singleQuery}{q}

\newcommand{\querySet}{Q}
\newcommand{\documentSet}{D}

\newcommand{\documentSpace}{\mathcal{\documentSet}}
\newcommand{\querySpace}{\mathcal{\querySet}}
\newcommand{\datasetSpace}{\mathcal{S}}
\newcommand{\parameterSpace}{\mathcal{W}}

\newcommand{\originalSet}{\ensuremath{\text{\textsf{\upshape{S}}}}}

\newcommand{\forgetSet}{\ensuremath{\text{\textsf{\upshape{F}}}}}
\newcommand{\entangledSet}{\ensuremath{\text{\textsf{\upshape{E}}}}}
\newcommand{\disjointSet}{\ensuremath{\text{\upshape{\textsf{D}}}}}
\newcommand{\retainSet}{\ensuremath{\text{\upshape{\textsf{R}}}}}

\newcommand{\learningAlgorithm}{M}
\newcommand{\unlearningAlgorithm}{U}

\newcommand{\initialModel}{\mathcal{M}_{\text{\upshape{init}}}}
\newcommand{\trainedModel}{\mathcal{M}_{\text{\upshape{train}}}}
\newcommand{\unlearnedModel}{\mathcal{M}_{\text{\upshape{unlearn}}}}

\newcommand{\retrainedModel}{\mathcal{M}_{\text{\upshape{retrain}}}}

\newcommand{\reals}{\mathbb{R}}

\newcommand{\relu}{\mathop{\mathrm{ReLu}}}

\newacronym{ir}{IR}{information retrieval}
\newacronym{drmm}{DRMM}{Deep Relevance Matching Model}
\newacronym{knrm}{KNRM}{Kernel-based Neural Ranking Model}
\newacronym{mrr}{MRR}{mean reciprocal rank}
\newacronym{marco}{MS MARCO}{MircoSoft MAchine Reading COmprehension}
\newacronym{trec}{TREC CAR}{TREC Complex Answer Retrieval}
\newacronym{nmu}{NuMuR}{Neural Machine UnRanking}
\newacronym{cocol}{CoCoL}{Contrastive and Consistent Loss}

\newcommand{\jerryRevise}[1]{{\textcolor{black}{#1}}}

\newglossaryentry{bertcat}
{
    name=BERTcat,
    description={BERT for concatenated query-document scoring}
}

\newglossaryentry{bertdot}
{
    name=BERTdot,
    description={BERT with Dot Production for separated query-document scoring}
}

\newglossaryentry{colbert}
{
    name=ColBERT,
    description={Contextualised Late Interaction over BERT}
}

\newglossaryentry{parade}
{
    name=PARADE,
    description={Passage representation aggregation for document reranking}
}

\newglossaryentry{duet}
{
    name=Duet,
    description={Duet architecture}
}
\newglossaryentry{matchp}
{
    name=MatchPyramid,
    description={Multi-level abstraction of interaction patterns between words, phrases and sentences, with a layer-by-layer architecture}
}

\begin{document}
\maketitle

\begin{abstract} 
We address the problem of machine unlearning in neural information retrieval (IR), introducing a novel task termed Neural Machine UnRanking (NuMuR). This problem is motivated by growing demands for data privacy compliance and selective information removal in neural IR systems. Existing task- or model- agnostic unlearning approaches, primarily designed for classification tasks, are suboptimal for NuMuR due to two core challenges: (1) neural rankers output unnormalised relevance scores rather than probability distributions, limiting the effectiveness of traditional teacher–student distillation frameworks; and (2) entangled data scenarios, where queries and documents appear simultaneously across both forget and retain sets, may degrade retention performance in existing methods.
To address these issues, we propose Contrastive and Consistent Loss (CoCoL), a dual-objective framework. CoCoL comprises (1) a contrastive loss that reduces relevance scores on forget sets while maintaining performance on entangled samples, and (2) a consistent loss that preserves accuracy on retain set.
Extensive experiments on MS MARCO and TREC CAR datasets, across four neural IR models, demonstrate that CoCoL achieves substantial forgetting with minimal retain and generalisation performance loss. Our method facilitates more effective and controllable data removal than existing techniques.

\end{abstract}


\section{Introduction } \label{sec:intro}
\glsresetall

Machine unlearning is the process of selectively removing specific data points from a trained machine-learning model~\citep{bourtoule2021machine, xu2023machine}. This task has gained significant attention in recent years as it addresses concerns regarding data privacy and model adaptability~\citep{VILLARONGA2018304, bourtoule2021machine, zhang2023review, xu2023machine}.

In this work, we focus on neural ranking models nowadays used for \emph{\gls{ir}}, i.e., on \emph{neural \gls{ir}}. In this context, machine unlearning may be needed for two main goals:
\begin{enumerate}[label = \alph*.]
    \item\label{enum:goals:a} \emph{addressing data-privacy concerns,} e.g., for deleting data of a user who has exercised their `right to be forgotten'~\citep{bourtoule2021machine,chen2021machine};
    \item\label{enum:goals:b} \emph{selectively deleting (e.g.\ outdated) information} \citep{teevan2013slow,campos2014survey}. For instance, an \gls{ir} system querying ``What are the EU member states?'' might need to exclude “UK” from its results post-2020~\cite{EuropeanCommission2024}, illustrating a practical application of machine unlearning in \gls{ir} systems.
\end{enumerate}

\begin{figure}[htbp]
    \centering
    \includegraphics[width=0.45\textwidth]{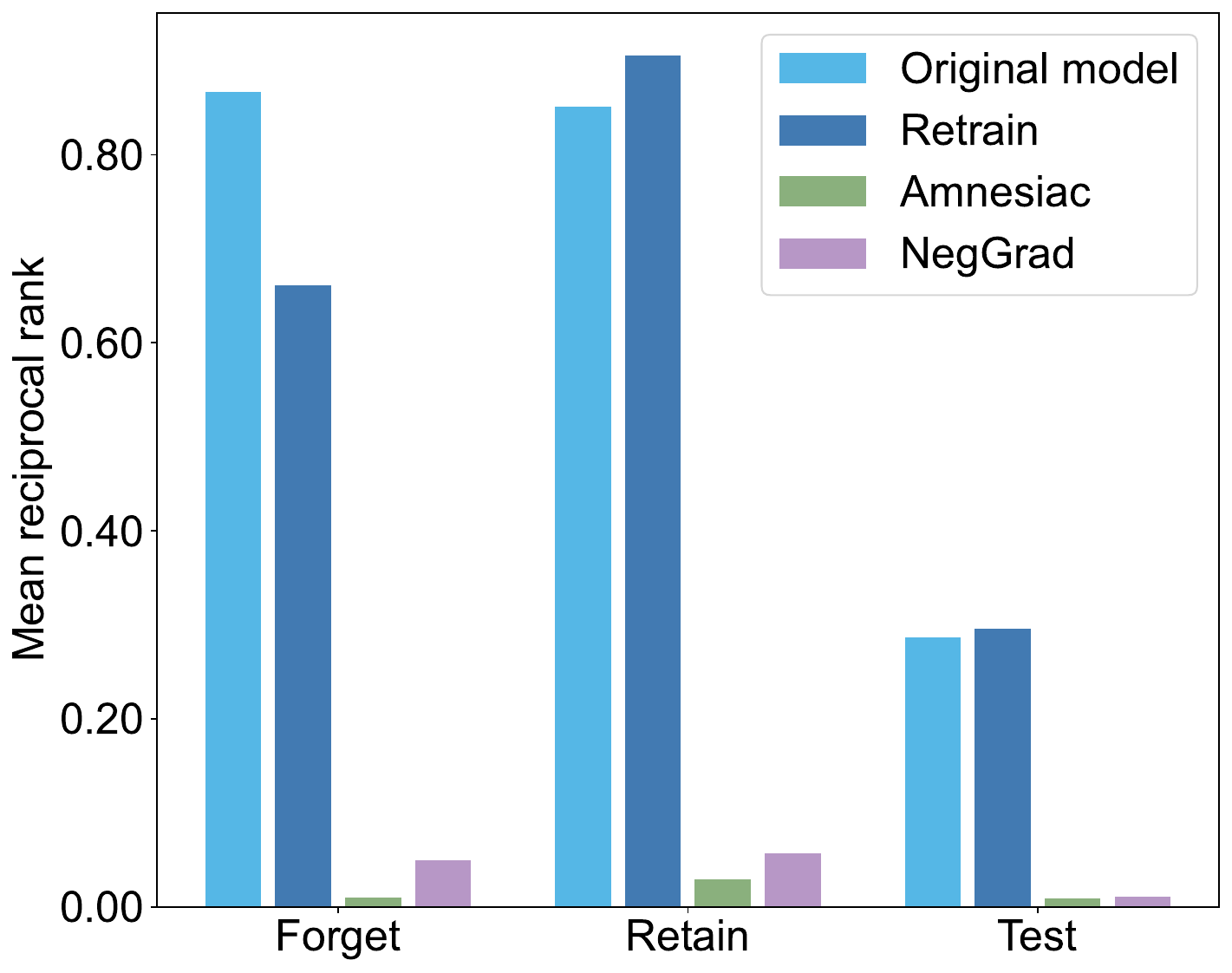}
    \caption{Breakdown (in the sense of performance degradation on retain and tests sets) of classical machine unlearning baselines in neural information retrieval. The retrieval model and dataset are \gls{colbert}~\cite{khattab2020colbert} and \gls{marco}~\cite{craswell2021ms}. }
    \label{figure:demo-result}
\end{figure}

It is therefore important to design methods for machine unlearning that can effectively deal with neural \gls{ir}. Prominent existing model- and task- agnostic unlearning methods like \emph{Amnesiac Unlearning}~\citep{Graves_Nagisetty_Ganesh_2021,Foster_Schoepf_Brintrup_2024} or \emph{Negative Gradient Removal}~\citep{Zhang2022,tarun2023fast} (NegGrad) could be employed. However, these have been primarily designed for classification scenarios where it is typically possible to unlearn a class by deliberately damaging the model accuracy on the samples within that class; and Figure~\ref{figure:demo-result} illustrates that such unlearning strategies perform poorly in neural \gls{ir} in the sense that reducing the performance of these models on the `forget set' (i.e.\ on the data to be removed) incurs a severe performance loss on the `retain set' (i.e.\ on the remaining data) and on test sets. We conjecture that this is due to strong dependencies in neural \gls{ir} models, where removing individual data points disrupts learned patterns~\cite{cao2015towards,Zhang2022,xu2024machine}.
Another model-agnostic unlearning method is the teacher--student framework~\citep{chundawat2023can,NEURIPS2023_062d711f} which was likewise originally designed for classification tasks.
However, as we discuss in detail in Section~\ref{subsec:breakdown}, a na\"ive application of this approach to neural \gls{ir} fails because the relevance scores generated by neural ranking models cannot easily be normalised.

An additional challenge is that unlearning solutions which perform well on Goal~\ref{enum:goals:a} may not be suitable for Goal~\ref{enum:goals:b} and vice versa. To see this, note that the ideal (though typically prohibitively costly) solution for Goal~\ref{enum:goals:a} would be to retrain the model from scratch on the retain set. However, even if such a `retrained' model was available, there is no guarantee that its performance on the forget set would be low enough to satisfy Goal~\ref{enum:goals:b}. Put differently, (re)training without the forget set does not achieve \emph{controllable forgetting,} i.e.\ the ability to regulate the degree of performance loss on the forget set whilst ensuring minimal loss in retention performance and inference capability.

In this work, we introduce machine unlearning methodology for both Goals~\ref{enum:goals:a} and \ref{enum:goals:b}. Our methodology is loosely based on the teacher--student framework from \citet{chundawat2023can} and \citet{NEURIPS2023_062d711f} but tailored to the challenges of neural \gls{ir}. Specifically, our contributions are as follows.

\begin{enumerate}
\item Formalisation of \emph{\gls{nmu}} -- the task of unlearning queries or documents within neural \gls{ir}, along with two datasets to benchmark \gls{nmu}.
\item Introduction of \emph{\gls{cocol}} -- a machine unlearning method specifically designed for the \gls{nmu} task.
\item Experimental validation demonstrating that \gls{cocol} improves upon baseline methods. Specifically, \gls{cocol} achieves controllable forgetting, enabling variable scales of data removal without markedly degrading the model's performance across both retain and test sets.
\end{enumerate}

\section{Background and problem definition} \label{sec:Background}
\glsreset{nmu}

In this section, we provide the first formal definition of machine unlearning within neural \gls{ir} and explain why existing teacher–student frameworks are inadequate for this domain.

\subsection{Machine unlearning} \label{subsec:machine_unlearning} 
Let $\parameterSpace \subseteq \reals^d$ be the \emph{parameter space} and let $\datasetSpace$ be the universe of possible datasets. Let $\learningAlgorithm \colon \datasetSpace \to \parameterSpace$ be a \emph{learning algorithm} which maps a \emph{training} set $\originalSet \in \datasetSpace$ to a \emph{model} $w \in \mathcal{W}$. Learning algorithms may be random but we do not make this explicit in the notation. The \emph{trained} model is then:
\begin{align}
  \trainedModel = M(\originalSet) \coloneqq \mathop{\mathrm{arg\,min}}_{w \in \parameterSpace} L_{ \originalSet}(w),
\end{align}
where $L_{ \originalSet}(w)$ is some suitable loss which typically penalises the discrepancy between the prediction by Model~$w$ and the ground truth contained in the data set $\originalSet$.

Given the training set $\originalSet$, let $\forgetSet \subseteq \originalSet$ be the \emph{forget set} which contains a subset of data points in $\originalSet$ to be unlearned; and let $\retainSet \coloneqq \originalSet \setminus \forgetSet \in \datasetSpace$ be the \emph{retain set} which contains the remaining data points. This defines the \emph{retrained} model
\begin{align}
  \retrainedModel\coloneqq M(\retainSet).
\end{align}
Let $\unlearningAlgorithm \colon \parameterSpace \times \datasetSpace \times \datasetSpace \to \parameterSpace$ be a (potentially random)  \emph{unlearning algorithm for $\learningAlgorithm$} which defines the \emph{unlearned} model
\begin{align}
  \unlearnedModel\coloneqq \unlearningAlgorithm(\trainedModel, \forgetSet, \retainSet).
\end{align}

Unlearning algorithms are normally expected to ensure that the unlearned model closely approximates the retrained model, i.e., $\unlearnedModel \approx \retrainedModel$, whilst the computational cost of unlearning -- starting from  $\trainedModel$ -- should be less than retraining from scratch on $\retainSet$ \citep{sekhari2021remember, chien2022efficient}. While mimicking $\retrainedModel$ aligns with Goal~\ref{enum:goals:a}, it does not enable controllable forgetting (Goal~\ref{enum:goals:b}). Therefore, we will base our unlearning approach on the \emph{teacher--student} framework (also known as \emph{knowledge distillation}) from \citet{chundawat2023can}, which can achieve pre-specified degrees of forgetting by implementing different distillation strategies. 

Informally, the teacher--student framework specifies the unlearning algorithm $\unlearningAlgorithm$ as using stochastic gradient-descent -- initialised from $\trainedModel$ -- to minimise (or at least decrease) 
\begin{align}
  L_{\mathcal{M}_{\forgetSet}, \forgetSet}(w) + L_{\mathcal{M}_\retainSet, \retainSet}(w), \label{eq:teacher_student_unlearning_objective}
\end{align}
where, for any dataset $\mathsf{A}$, the objective $L_{\mathcal{M}, \mathsf{A}}(w)$ penalises the difference between predictions made by the \emph{student} model $w \in \parameterSpace$ (which is unlearning) and some fixed \emph{teacher} model $\mathcal{M}$ on $\mathsf{A}$ and is typically specified as follows.

\begin{itemize}
    \item Since $\unlearnedModel$ should perform similarly to $\retrainedModel$ on $\retainSet$ which in turn should perform similar to $\trainedModel$ (on $\retainSet$), it is common to take $\mathcal{M}_\retainSet \coloneqq \trainedModel$ in \eqref{eq:teacher_student_unlearning_objective}.
   This can be interpreted as training $w$ to obey the `competent' teacher model $\trainedModel$ on $\retainSet$ \citep{kim2022efficient,10097553,chundawat2023can,NEURIPS2023_062d711f}.
   
    \item Since $\unlearnedModel$ should achieve controllable forgetting, i.e., achieve a pre-specified performance $\delta$ that is worse than $\trainedModel$ on $\forgetSet$, it is common to take
    \begin{align}
      L_{\mathcal{M}_\forgetSet, \forgetSet}
      & \coloneqq
      - L_{\trainedModel, \forgetSet},
    \end{align}
    in \eqref{eq:teacher_student_unlearning_objective} which can be viewed as training $w$ to disobey the `competent' teacher $\trainedModel$ on $\forgetSet$ \citep{NEURIPS2023_062d711f}; and then to stop the gradient-descent iterations when the accuracy on the forget set has dropped to the target level $\delta$. Alternatively, if the goals is that the unlearned model should perform similarly to $ \initialModel \coloneqq M(\emptyset)$ on $\forgetSet$, one could simply take $\mathcal{M}_\forgetSet \coloneqq \initialModel$ in \eqref{eq:teacher_student_unlearning_objective}, which can be viewed as training $w$ to obey the `incompetent' teacher model $\initialModel$ on $\forgetSet$ \citep{chundawat2023can}. Of course, $\initialModel$ could be replaced by another model, e.g., by an adversarial model trained on with noisy data \cite{kim2022efficient, tarun2023fast, 10097553}.
\end{itemize}

\subsection{Unlearning in neural information retrieval}
\glsreset{ir}
The goal of \emph{\gls{ir}} is to identify and retrieve documents in response to a search query \cite{Ceri2013}. Let $\querySpace$ be the universe of potential queries and let $\documentSpace$ be the universe of potential documents. Queries are user inputs or requests for specific information, typically in the form of words, phrases, or questions; documents refer to units of content, such as web pages or articles.

Then a dataset for (neural) \gls{ir} $\originalSet \in \datasetSpace$ consists of tuples $(x, y)$, where
\begin{itemize}
 \item $x = \langle\singleQuery, \singleDocument\rangle \in \querySpace \times \documentSpace$ is a query--document pair;
 \item $y \in \{\text{+}, \text{-}\}$ is the ground-truth relevance label of $\langle\singleQuery, \singleDocument\rangle$. Here, `+' indicates that $d$ is considered relevant to $q$; `-' indicates that $d$ is irrelevant to $q$.
\end{itemize}

A \emph{neural-ranking model} $w \in \parameterSpace$ is then trained to predict a relevance score $f_w(x) \in \reals$ of some query--document pair $x = \langle \singleQuery, \singleDocument\rangle$. Relevance scores output by neural-ranking models are used to rank documents. Each document associated with a query is sorted by its score in (descending) order so that higher scores correspond to a higher rank and thus earlier positions in the search results. 

\emph{\Gls{nmu}} is then the task that this model unlearns either queries or documents (or both):
\begin{itemize}
    \item \emph{Query removal} refers to deleting a set of queries $\querySet'$ (and associated relevance scores) from the dataset. In this case, $\forgetSet \coloneqq \{(\langle \singleQuery, \singleDocument\rangle, y) \in \originalSet \mid \singleQuery \in \querySet'\}$.
    
    \item \emph{Document removal} refers to deleting a set of documents $\documentSet'$ (and associated relevance scores) from the dataset. In this case, $\forgetSet \coloneqq \{(\langle \singleQuery, \singleDocument\rangle, y) \in \originalSet \mid \singleDocument \in \documentSet'\}$.
\end{itemize}

One of the difficulties encountered in \gls{nmu} is that certain queries or documents may appear simultaneously in the retain set $\retainSet$ and in the forget set $\forgetSet$. For example, assume that the query: ``The best one-week itinerary for a trip to London'' is associated with two recommended itineraries (i.e., documents). Consider if one itinerary's owner recalls their answer, we must unlearn one query--document pair whilst maintaining the other.

\begin{figure*}[htbp]
\begin{subfigure}{.48\textwidth}
    \centering
    \includegraphics[scale=0.7]{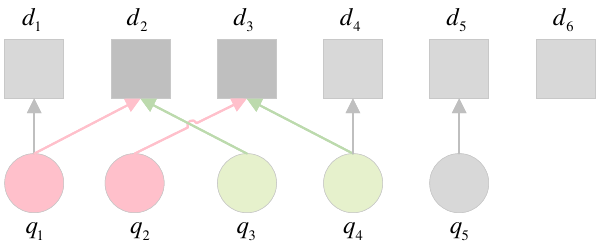}
    \caption{Query removal.}
    \label{figure.query.removal.demo}
    \end{subfigure}
    \hfill 
\begin{subfigure}{.48\textwidth}
    \centering
    \includegraphics[scale=0.7]{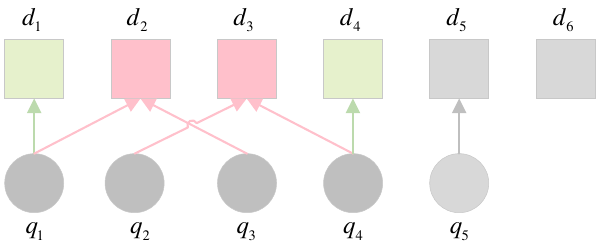}
    \caption{Document removal.}\label{figure.document.removal.demo}
    \end{subfigure}
  \caption{\raggedright Illustration of machine unranking for the dataset. $\originalSet = \{(\langle\singleQuery_1, \singleDocument_1\rangle, y_{1,1}), (\langle\singleQuery_1, \singleDocument_2\rangle, y_{1,2}), (\langle\singleQuery_2, \singleDocument_2\rangle, y_{2,2}), (\langle\singleQuery_3, \singleDocument_2\rangle, y_{2,2}), \allowbreak(\langle\singleQuery_4, \singleDocument_3\rangle, y_{4,3}), (\langle\singleQuery_4, \singleDocument_4\rangle, y_{4,4}), (\langle\singleQuery_5, \singleDocument_5\rangle, y_{5,5}) \}$. \quad \quad
  (a) Query removal. Here, queries $Q' = \{q_1, q_2\}$ are to be unlearned. Thus, $\forgetSet = \{(\langle\singleQuery_1, \singleDocument_1\rangle, y_{1,1}), (\langle\singleQuery_1, \singleDocument_2\rangle, y_{1,2}), (\langle\singleQuery_2, \singleDocument_2\rangle, y_{2,2})\}$, $\entangledSet = \{(\langle\singleQuery_3, \singleDocument_2\rangle, y_{2,2}), (\langle\singleQuery_4, \singleDocument_3\rangle, y_{4,3})\}$ and $\disjointSet = \{(\langle\singleQuery_4, \singleDocument_4\rangle, y_{4,4}), (\langle\singleQuery_5, \singleDocument_5\rangle, y_{5,5})\}$.
   \quad \quad (b) Document removal. Here, documents $D'=\{d_2, d_3\}$ are to be unlearned. Thus, $\forgetSet = \{(\langle\singleQuery_1, \singleDocument_2\rangle, y_{1,2}), (\langle\singleQuery_2, \singleDocument_3\rangle, y_{2,3}), (\langle\singleQuery_3, \singleDocument_2\rangle, y_{3,2}), (\langle\singleQuery_4, \singleDocument_3\rangle, y_{4,3})\}$, $\entangledSet = \{(\langle\singleQuery_1, \singleDocument_1\rangle, y_{1,1}), (\langle\singleQuery_4, \singleDocument_4\rangle, y_{4,4})\}$ and $\disjointSet = \{(\langle\singleQuery_5, \singleDocument_5\rangle, y_{5,5})\}$.}
  \label{figure:unranking-demo}
\end{figure*}

To formalise this issue, we split the retain set $\retainSet$ into an \emph{entangled} set $\entangledSet \coloneqq \{(\langle \singleQuery, \singleDocument\rangle, y) \in \retainSet \mid  
  \exists \, (\langle \singleQuery', \singleDocument'\rangle, y) \in \forgetSet: (\singleQuery \in Q' \text{ or } \singleDocument \in D')\}$ containing queries or documents that also appear in the forget set and a \emph{disjoint set} $\disjointSet \coloneqq \retainSet \setminus \entangledSet$ containing all other queries and documents.
Here, $Q'$ and $D'$ are again sets of queries and documents that should be unlearned. Figure~\ref{figure:unranking-demo} provides detailed illustrations of entangled sets and disjoint sets in both query removal and document removal. 

\subsection{Breakdown of existing knowledge-distillation based unlearning algorithms in neural information retrieval} 
\label{subsec:breakdown}

The teacher--student approach from \citet{chundawat2023can} (reviewed in Section~\ref{subsec:machine_unlearning}) was primarily designed for classification models where the objectives $L_{\mathcal{M}, \mathsf{A}}(w)$ in \eqref{eq:teacher_student_unlearning_objective} penalise the difference between class probabilities predicted by the student model $w$ and by the reference (`teacher') model $\mathcal{M}$. The teacher--student approach thus exploits the fact that the outputs of neural classification models are class probabilities which are always normalised, so that forgetting (e.g., of a class) can always be ensured by simply lowering the associated class probabilities. However, the relevance scores generated by neural ranking models cannot typically be normalised so that the ranking implications of modifying relevance-score distributions are unclear (see, e.g., Figure~\ref{figure:compare_train_init} in Appendix~\ref{sec:re.comparison}); and this, as well as the fact that classification tasks do not involve entangled sets, causes the teacher--student approach to break down in neural \gls{ir}. More specifically, we identify the following problems:
\begin{enumerate}
    \item \emph{$\initialModel$ cannot serve as `incompetent' teacher.} Due to the lack of normalisation, the relevance scores on $\forgetSet$ are not necessarily lower under the `incompetent' teacher model (e.g., $\initialModel$) than under the trained model. For example, a relevance score of 30, say, might imply a high rank under the `competent' teacher model, but a low rank under the incompetent teacher model (see Figure~\ref{figure:compare_train_init} in Appendix~\ref{sec:re.comparison}). Therefore, using $\initialModel$ as the teacher for the forget set while employing $\trainedModel$ as a `competent' teacher for the retain set may be counterproductive.
    
    \item \emph{Controllable forgetting is challenging in neural ranking.} In a $k$-class classification model, forgetting a specific sample is achieved by adjusting the model output (i.e., class probabilities $\in \reals^k$) so that the probability of the correct class falls below $1/k$. This type of forgetting can be quantitatively assessed using the Kullback--Leibler divergence from the outputs of the teacher model to the student model \cite{kim2022efficient,NEURIPS2023_062d711f,10097553}. However, in neural ranking models, achieving forgetting by manipulating the model output (i.e., relevance score $\in \reals$, an unnormalised scalar) is challenging, due to the absence of a clear threshold or benchmark for adjusting these scores.

    \item \emph{A na\"ive application of the teacher--student framework overlooks the entangled set.} 
     A challenge in \gls{nmu} is that some queries or documents can appear in both the retain and the forget set, as formalised by the entangled set.  Conventional teacher--student frameworks implement distinct strategies for the forget and retain sets, as summarised in \eqref{eq:teacher_student_unlearning_objective}, without accounting for the entangled set. However, effectively decoupling the learned relevance estimation patterns between the forget set and the entangled set using such unlearning methods is problematic. This issue will be evidenced in Figure~\ref{figure:ablation1}, where we illustrate that teacher--student approach that ignores the entangled set yields inferior performance on both the retain and test sets. 
\end{enumerate}

\section{Proposed neural machine unranking methodology}  \label{sec:method}

\glsreset{cocol}
In this section, we propose a new teacher--student framework for \gls{nmu}, called \emph{\gls{cocol}}. To address the three challenges discussed at the end of Section~\ref{sec:Background}, \gls{cocol} introduces the following elements.

\begin{enumerate}
  \item To overcome the problem of using an `incompetent' teacher model (such as $\initialModel$) in the presence of unnormalised relevance scores, we attempt to reduce the relevance scores on the forget set (relative to the trained model, $\trainedModel$) whilst seeking to maintain the relevance scores on the retain set.
  \item Given that relevance scores are not normalised, to enable controllable forgetting, we stop the unlearning iterations when
 \begin{equation}
    \frac{1}{\# Q_\forgetSet} \sum_{q \in Q_\forgetSet} \frac{1}{\mathrm{rank}_{w}(q)}, \label{eq:mrr}
 \end{equation}
 is approximately equal to some pre-specified target $\delta > 0$  rather than basing the termination on the average relevance score reaching a predefined target level. Here,  $Q_\forgetSet \coloneqq \{q \in \querySpace \mid \exists \, (\langle q',d\rangle, y) \in \forgetSet: q' = q\}$ is the set of distinct queries in the forget set. 
 \glsreset{mrr}
\begin{itemize}
\item In \emph{query removal}, $\mathrm{rank}_w(q)$ denotes the rank of the first relevant document for query $q$ among all documents allocated to query $q$ for ranking. evaluated by Model~$w$. Here, \eqref{eq:mrr} simplifies to the classical \gls{mrr} as described by \citet{liu2009learning}.
\item In \emph{document removal}, $\mathrm{rank}_w(q)$ represents the rank of the first document marked for removal. This may differ from the rank of the first relevant document. For example, if Model~$w$ ranks the documents for Query~$q$ as $[d_1, d_3, d_4, d_2, \ldots]$, where $d_1$ is the first relevant but $d_2$ is the first marked for removal, the reciprocal rank is recalculated as $0.25$. While this differs from the classical \gls{mrr}, we retain the `MRR' notation for consistency in evaluation metrics.
\end{itemize}

\item To ensure that reducing the model accuracy on the forget set does not inadvertently damage the model performance on the entangled set, we pair a `forgetting sample' with a random selection of a sample from the corresponding entangled set, as explained in the next section.
\end{enumerate}

\subsection{Objective}


\gls{cocol} uses gradient steps, started from the trained model $\trainedModel$, to decrease an objective of the form 
\begin{align}
    L_{\mathcal{M}_{\forgetSet \cup \entangledSet} \forgetSet \cup \entangledSet}(w) + L_{\mathcal{M}_\disjointSet, \disjointSet}(w),
    \label{eq:unrank}
\end{align}
were $L_{\mathcal{M}, \mathsf{A}}(w)$ is again some objective which penalises the discrepancy between $w$ and some reference model $\mathcal{M}$ on some dataset $\mathsf{A}$. Note that this objective differs from the standard teacher--student framework \eqref{eq:teacher_student_unlearning_objective} in that the entangled set is moved into the first component. Note also that we say `decrease' rather than `minimise' because the unlearning is simply stopped when a pre-defined level of forgetting has been achieved (see below for details).

The components $L_{\mathcal{M}_{\forgetSet \cup \entangledSet} \forgetSet \cup \entangledSet}(w)$ and $L_{\mathcal{M}_\disjointSet, \disjointSet}(w)$ are implicitly defined through update rules which we now specify, where for some query--document pair $x = \langle q, d\rangle$:

\begin{equation}
\jerryRevise{
    \Delta_{w, \mathcal{M}}(x) = \frac{f_\mathcal{M}(x) - f_w(x) }{f_\mathcal{M}(x) + f_w(x)},  \label{eq:performance_delta}}
\end{equation}
\jerryRevise{measures the discrepancy between the relevance score $f_\mathcal{M}(x)$ returned by some fixed reference (`teacher') model $\mathcal{M}$ and the relevance score $f_w(x)$ returned by the `student' model $w$.  This measure supports maintaining relevance consistency for.}

\jerryRevise{To facilitate targeted forgetting, for each $(x, y) = (\langle q, d \rangle, y) \in \forgetSet$, we define the query-specific set of samples in $\originalSet$:}

\begin{align}
\jerryRevise{
 \originalSet_{x} =\originalSet_{\langle \singleQuery, \singleDocument \rangle} \coloneqq \{ \langle q^*, d^* \rangle \in \mathcal{Q} \times \mathcal{D} \mid \exists \left( \langle q^*, d^* \rangle, y^* \right) \in \originalSet} : \\ \jerryRevise{q^* =q \}}
\end{align}
\jerryRevise{and compute the minimal relevance score for query~$q$ in the teacher model:}
\begin{equation}
\jerryRevise{
s^{\text{min}}_\mathcal{M}(x) = \min_{x \in \originalSet_{x}} f_\mathcal{M}(x)
\label{eq:min_score_teacher}}
\end{equation}

\jerryRevise{We then define an adjusted discrepancy measure to suppress the relevance score for forget set:}
\begin{equation}
\jerryRevise{
     \Delta^{\text{min}}_{w, \mathcal{M}}(x)  = -\frac{s^{\text{min}}_\mathcal{M}(x) - f_w(x) }{s^{\text{min}}_\mathcal{M}(x) + f_w(x)},  \label{eq:performance_delta_adjusted}}
\end{equation}
\jerryRevise{Here, \eqref{eq:performance_delta_adjusted} prioritises reducing $f_w(x)$ toward the minimal relevance score $s^{\min}_\mathcal{M}(x)$ produced by $\mathcal{M}$ for query $q$.}

The update rules are then as follows.
\begin{enumerate}
    \item \textit{Contrastive loss: implicit definition of $L_{\mathcal{M}_{\forgetSet \cup \entangledSet} \forgetSet \cup \entangledSet}(w)$.} We employ a \emph{contrastive} loss to modify the student model $w$ such that it generates lower relevance scores on the forget set than the trained model $\trainedModel$ whilst ensuring that the relevance scores on the entangled set are maintained. Specifically, at each iteration, we randomly select a sample $(x, y) = (\langle q, d\rangle, y) \in \forgetSet$ from the forget set and a second sample $(x', y') = (\langle q', d'\rangle, y') \in \entangledSet_{\langle q, d\rangle}$, where $\entangledSet_{\langle q, d\rangle} \coloneqq \{(\langle q'', d''\rangle, y'') \in \entangledSet \mid q'' = q \text{ or } d'' = d\}$ contains the samples that are entangled with $(x, y)$, and then take a gradient step which reduces

\begin{equation}
    \jerryRevise{
    \mathrm{ReLu}(\Delta^\text{min}_{w, \trainedModel}(x)) + \lvert \Delta_{w, \trainedModel}(x')\rvert. \label{eq:contrastive_loss}}
\end{equation}%

Here, $\relu(z) \coloneqq \text{max}(0,z)$. If $\entangledSet_{\langle q, d\rangle} = \emptyset$ then we take the second term in \eqref{eq:contrastive_loss} to be zero.

    \item \textit{Consistent loss: implicit definition of $L_{\mathcal{M}_\disjointSet, \disjointSet}(w)$.}  We employ a \emph{consistent} loss to modify the student model $w$ such that it generates relevance scores  on the disjoint set that are similar to those from the trained model $\trainedModel$. Specifically, at each iteration, we randomly select a positive (i.e., relevant) sample  $(x^+, y^+) = (\langle q^+, d^+\rangle, y^+) \in \disjointSet$ and a negative (i.e., irrelevant) sample $(x^-, y^-) = (\langle q^-, d^-\rangle, y^-) \in \disjointSet$ from the disjoint set and then take a gradient step which reduces

\begin{equation}
     \jerryRevise{
    \lvert \Delta_{w, \trainedModel}(x^+)\rvert + \lvert \Delta_{w, \trainedModel}(x^-)\rvert.  \label{eq:consistent_loss}}
\end{equation}
\end{enumerate}

\begin{figure}[htbp]
    \centering
    \includegraphics[width=0.5\textwidth]{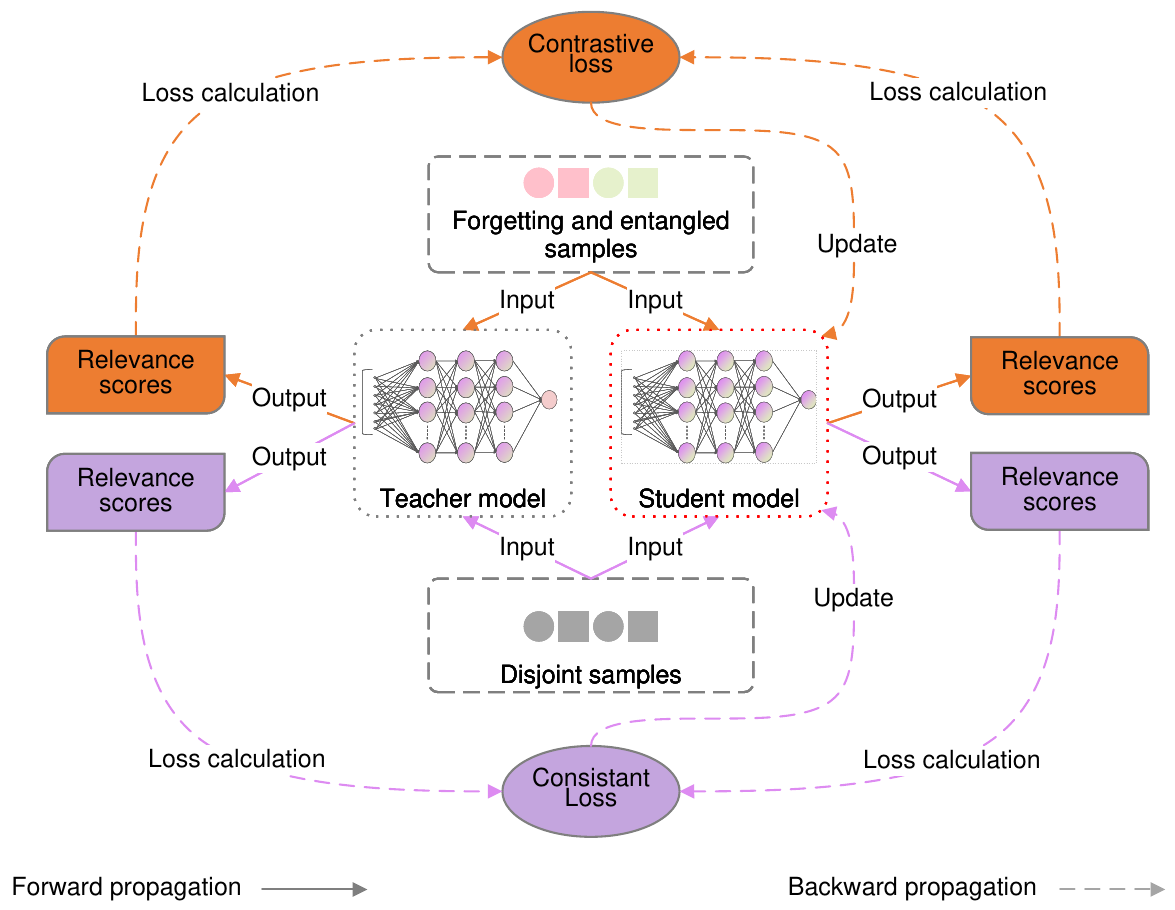}
    \caption{Illustration of the proposed \gls{cocol} method.}
    \label{figure:method-illustration}
\end{figure}

In summary, our \gls{cocol} unranking approach is illustrated in Figure~\ref{figure:method-illustration}.

\subsection{Stopping criteria of unranking}
\begin{algorithm}[htbp]
    \footnotesize
    \caption{\gls{nmu} procedure based on \gls{cocol}}
    \label{alg:cocol}
    \begin{algorithmic}[1]
        \Require Trained model $\trainedModel$, forget set $\forgetSet$, entangled set $\entangledSet$, disjoint set $\disjointSet$, target forgetting level $\delta$, maximum epochs $N$
        \Ensure Updated model $w^*$
        
        \State Set model parameters $w_0 \gets \trainedModel$
        \For{$e = 1$ to $N$}
            \State // Phase 1: Forget and entangled set updates
            \ForAll{$(x, y) \in \forgetSet \cup \entangledSet$}
                \State Compute loss $\mathcal{L}_\text{forget}(w_e)$ using Equation~\eqref{eq:contrastive_loss}
                \State Update $w_e$ by gradient step on $\mathcal{L}_\text{forget}$
            \EndFor
            
            \State // Phase 2: Disjoint set updates
            \ForAll{$(x, y) \in \disjointSet$}
                \State Compute loss $\mathcal{L}_\text{retain}(w_e)$ using Equation~\eqref{eq:consistent_loss}
                \State Update $w_e$ by gradient step on $\mathcal{L}_\text{retain}$
            \EndFor
            
            \State Compute $\mathrm{MRR}_{\forgetSet}(w_e)$
            
            \If{$\mathrm{MRR}_{\forgetSet}(w_e) \leq \delta$}
                \State \textbf{break} // Early stopping
            \EndIf
        \EndFor
        \State \Return Final model $w^* \gets w_e$
    \end{algorithmic}
\end{algorithm}

\jerryRevise{As shown in algorithm~\ref{alg:cocol}, theocratically, alternating between \eqref{eq:contrastive_loss} and \eqref{eq:consistent_loss} ensures stable performance on both entangled and disjoint datasets, while performance on the forget set will progressively decline. Therefore, the optimal time to stop unlearning is when the performance on the forget set as measured by \eqref{eq:mrr} reaches the pre-specified level $\delta > 0$.}

\section{Experiments} \label{sec:exp}
\subsection{Datasets}

Currently, there are no existing \gls{ir} datasets specifically designed for machine unlearning research. To address this, we propose curating datasets derived from established benchmark \gls{ir} datasets. In selecting the appropriate datasets for \gls{nmu}, our selection criteria focused on datasets that feature extensive one-to-many relevant query--document and document--query pairings, essential for evaluating \gls{nmu} methodologies. An in-depth review of resources listed on \href{https://ir-datasets.com/}{ir-datasets.com} identified two sources that fulfil these requirements: \gls{marco} \citep{craswell2021ms} and \gls{trec} \citep{dietz2019trec}. These sources were selected due to their large sample sizes and the presence of overlapping queries and documents. Table~\ref{table.dataset} summarises the datasets.

\begin{table}[htbp]
\caption{Datasets created for this study.}
\centering
\resizebox{0.5\textwidth}{!}{
\begin{tabular}{llrr}
\toprule
Task & Item & \!\!\!\!\!\!\!\!\!\!\!\!\!\!\!\!\!\!\!\!\gls{marco} & \gls{trec} \\\midrule
\multirow{5}{*}{\makecell[l]{Document\\Removal}} & Queries with multiple positive documents\!\!\!\!\!\!\!\!\!\!\!\!\!\!\!\!\!\!\!\! & 2782 & 220271\\
 & Positive documents per query\!\!\!\!\!\!\!\!\!\! & 2.19 & 3.42\\
 & To-be-ranked documents per query\!\!\!\!\!\!\!\!\!\! & 2153 & 100\\
 & Pairwise samples for training & 5983761 & 1945509\\
 & Pairwise samples for test & 6668967 & 4710706\\ \midrule
\multirow{5}{*}{\makecell[l]{Query\\Removal}} & Positive passages with multiple queries\!\!\!\!\!\!\!\!\!\! & 4035 & 19455 \\
 & Associated queries per positive documents\!\!\!\!\!\!\!\! & 2.1 & 3.33\\
 & To-be-ranked documents per query\!\!\!\!\!\!\!\!\!\! & 1986 & 100 \\
 & Pairwise samples for training & 8005618 & 752003\\
 & Pairwise samples for test & 6668967 & 4710706 \\ \midrule
\end{tabular}%
}
\label{table.dataset}
\end{table}

\subsection{Evaluation metrics}

\subsubsection{Unlearning performance} \label{sec:unlearn.metric}

To evaluate ranking performance, we use the \emph{\gls{mrr}} metric as defined in \eqref{eq:mrr} on the forget set. For the retain and test sets, \gls{mrr} is similarly computed as the average of the reciprocals of the rank positions of the first retrieved relevant document for the queries in each set.

\subsubsection{Unlearning time} 
To ensure consistent measurement across different neural ranking models and unlearning methods we report the \emph{normalised unlearn epoch duration:} 
\begin{align}
  \MoveEqLeft \text{(normalised unlearn epoch duration)}\\
  & \coloneqq 
  \frac{\text{(avg time per unlearn epoch of $\unlearnedModel$)}}{\text{(avg time per learning epoch of $\trainedModel$)}},
\end{align}
as well as the \emph{total unlearn time:}
\begin{align}
  \text{(total unlearn time)}
  & \coloneqq 
  \text{(normalised unlearn epoch duration)} \\
  & \qquad \times \text{(no of unlearn epochs)}.
\end{align}

\subsection{NIR models and unlearning baselines}

\jerryRevise{We evaluate the proposed \gls{cocol} framework on four neural ranking models for \gls{ir} that leverage pretrained language models. Specifically, we consider \emph{\gls{bertcat}}~\cite{hofstatter2021efficiently}, which jointly encodes concatenated query-document pairs for relevance estimation; \emph{\gls{bertdot}}~\cite{hofstatter2021efficiently}, which encodes queries and documents separately and computes relevance via dot-product similarity; \emph{\gls{colbert}}~\cite{khattab2020colbert}, which employs contextualised late interaction to balance expressiveness and efficiency; and \emph{\gls{parade}}~\cite{li2023parade}, which aggregates passage-level representations for improved document reranking.}

Given the limited studies that exist in \gls{nmu}, identifying comparable baselines is challenging. Therefore, the following task- and model-agnostic unlearning methods were selected as baselines: 
\begin{enumerate}
    \item \jerryRevise{\emph{Catastrophic Forgetting (CF)} \citep{goel2024corrective}: As a straightforward strategy, CF continues training $\trainedModel$ exclusively on the retain set, relying on the natural forgetting phenomenon in neural networks \citep{kirkpatrick2017overcoming} to erase knowledge of the forget set.}
    
    \item \emph{Amnesiac} \cite{Graves_Nagisetty_Ganesh_2021,Foster_Schoepf_Brintrup_2024} continues training on $\trainedModel$ but with mislabelled samples in the forget set\footnote{\citet{Graves_Nagisetty_Ganesh_2021} proposed two unlearning methods: the first method is as described, while the second requires gradient storage during the training of $\trainedModel$ and is challenging to apply in neural ranking tasks. Following \citet{Foster_Schoepf_Brintrup_2024}, we use only the first method and refer to it as `Amnesiac'.
    }. To adapt this idea to \gls{nmu}, we intentionally score several $\langle \singleQuery,\singleDocument \rangle$ pairs marked as `negative' higher than those labelled as `positive' in the forget set and then keep training $\trainedModel$ on the revised forget set and the original entangled set.

    \item \emph{NegGrad}, short for `negative gradient', updates a learned model in the reverse direction of the original gradient on forget-set samples~\cite{Zhang2022,tarun2023fast}.
    
    \item \jerryRevise{\emph{Selective Synaptic Dampening (SSD)}\citep{Foster_Schoepf_Brintrup_2024}: SSD edits model parameters directly based on the relative importance of forget and retain samples to $\trainedModel$.}
    
    \item \jerryRevise{\emph{Bad Teacher (BadT)} \citep{chundawat2023can}: This method employs a randomly initialised model as an intentionally incompetent teacher on the forget set, while retaining $\trainedModel$ as a competent teacher for the retained set. This setup discourages the student model from preserving knowledge about the forget set during distillation.}
\end{enumerate}

We also report results for \emph{retrain}, i.e., for retraining from scratch ($\initialModel$) on only the retain set \cite{sekhari2021remember,bourtoule2021machine,chien2022efficient,xu2023machine} as this can be considered the `idealised' approach (unless `controllable forgetting' is sought). However, recall that as explained in Section~\ref{subsec:machine_unlearning}, obtaining $\retrainedModel$ is typically prohibitively costly.

\begin{figure*}[htbp]
    \centering
    \begin{subfigure}[t]{0.3\textwidth}
        \centering
        \includegraphics[width=\textwidth]{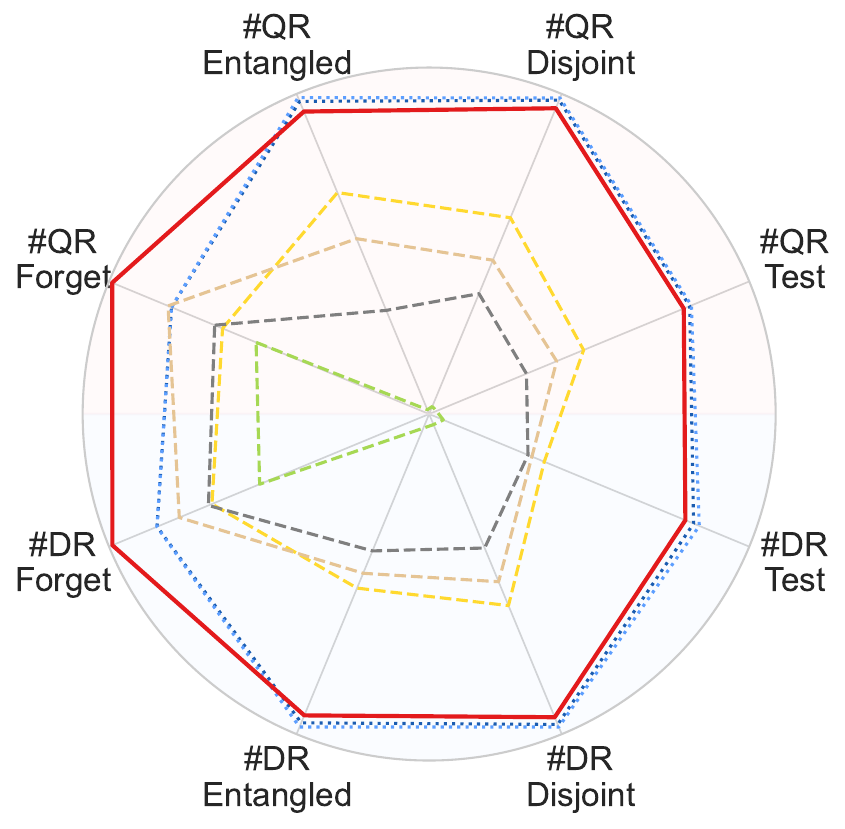}
        \caption{5\% Removal}
        \label{fig:avg_5}
    \end{subfigure}
    \hfill
    \begin{subfigure}[t]{0.3\textwidth}
        \centering
        \includegraphics[width=\textwidth]{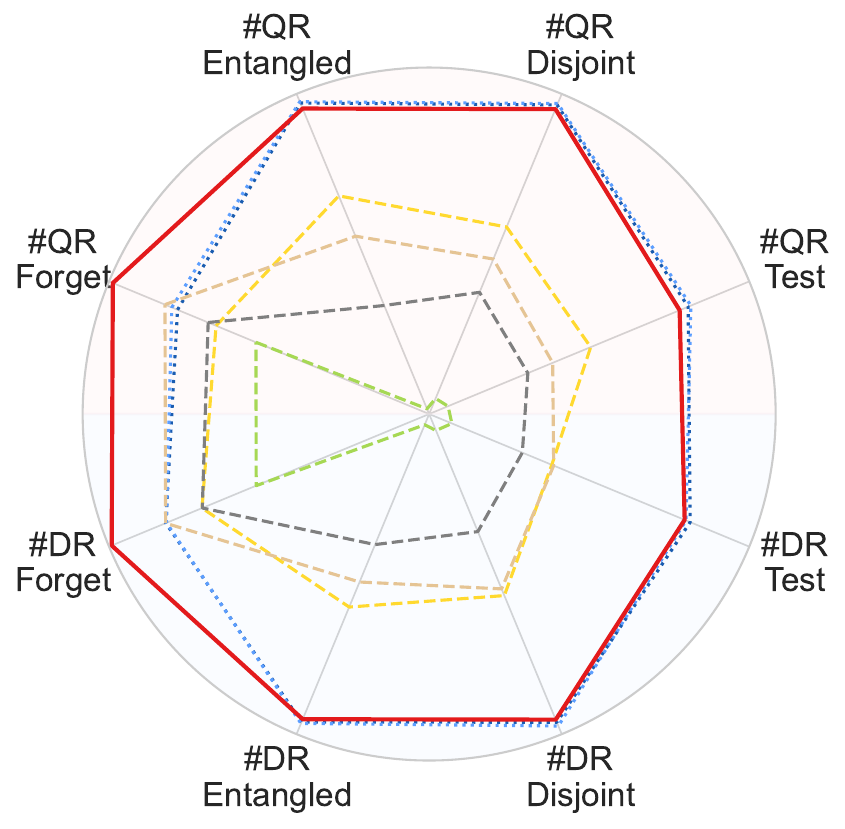}
        \caption{15\% Removal}
        \label{fig:avg_15}
    \end{subfigure}
    \hfill
    \begin{subfigure}[t]{0.3\textwidth}
        \centering
        \includegraphics[width=\textwidth]{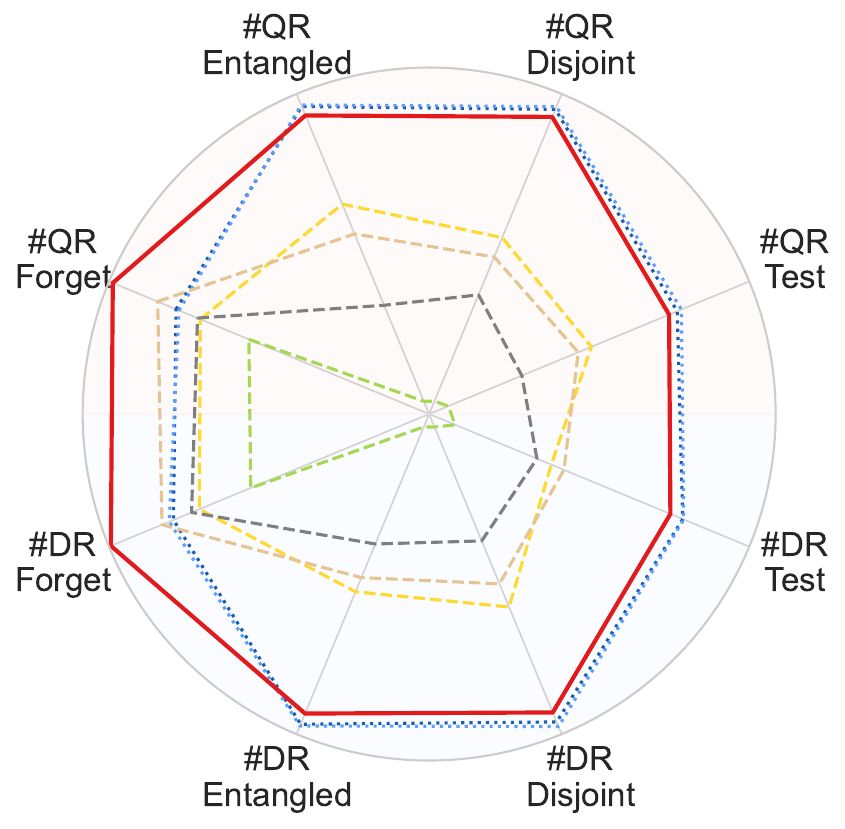}
        \caption{25\% Removal}
        \label{fig:avg_25}
    \end{subfigure}

    \vspace{0.1em}
    \begin{adjustbox}{center}
        \includegraphics[width=0.85\textwidth]{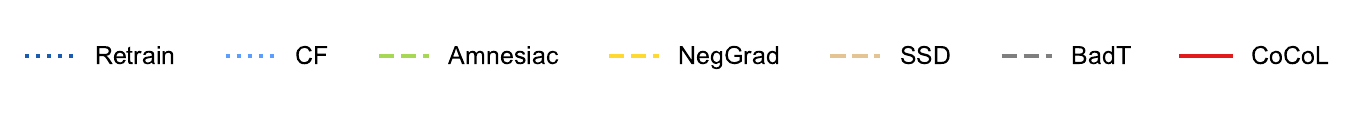}
    \end{adjustbox}

\caption{Average unlearning performance across datasets and models for varying forget set sizes. \#QR and \#DR denote query removal and document removal, respectively. To better visualise forgetting ability, the forget set performance is normalised as $1 - |\text{MRR}^{\unlearnedModel}_{\text{forget}} - \text{MRR}^{\retrainedModel}_{\text{test}}|$, where $\text{MRR}_{\retrainedModel}^{\text{test}}$ denotes the MRR of $\retrainedModel$ on the test set. Higher values indicate closer alignment with $\retrainedModel$ (ideal performance on the vertex). The Y-axis ranges from 0 to 1 for all sets except the test set, which ranges from 0 to 0.5.}
    \label{fig:average_radar}
\end{figure*}

\subsection{Experimental result}

\subsubsection{Unlearning performance under varying forget set sizes}

First, aligning with Goal~\ref{enum:goals:a}, we set $\eqref{eq:mrr}$ to match the test performance of the $\retrainedModel$.

To evaluate the impact of varying forget set sizes, we conduct three groups of experiments where 5\%, 15\%, and 25\% of query-document pairs are designated for forgetting, corresponding to their proportion relative to the total relevant query-document pairs.

Figure~\ref{fig:average_radar} summarises the results. For clarity, we report the average performance of different unlearning methods across both datasets and all four neural \gls{ir} models. Detailed results for each experimental setting are provided in Table~\ref{tab:detail_maro} and Table~\ref{tab:detail_trec} in the Appendix.

On the forget set, consistent with Goal~\ref{enum:goals:a}, \gls{cocol} achieves performance metrics that are most aligned with those of test set performance of $\retrainedModel$. As depicted in Figure~\ref{fig:average_radar}, \gls{cocol}'s performance under both query removal and document removal settings remains closest to the ideal vertex, outperforming all baselines in maintaining this alignment. Notably, even the `gold standard' Retrain \citep{goel2024corrective}  exhibits discrepancies between its performance on the forget set and the test set, a phenomenon observed across all three forget set sizes.
For the entangled and disjoint sets, ideal unlearning requires performance comparable to $\retrainedModel$. All baseline methods except CF consistently exhibit performance degradation on these sets relative to the benchmark. In contrast, \gls{cocol}, maintains performance close to  $\retrainedModel$ across both sets. This stability highlights its capacity for effective forgetting.

Regarding the test set (unseen data), Retrain serves as the benchmark, where higher scores indicate better generalisation. Apart from CF, all baselines suffer notable drops in \gls{mrr}, with Amnesiac exhibiting the most severe degradation. In contrast, \gls{cocol} not only achieves superior forgetting but also sustains robust generalisation, underscoring its effectiveness in preserving inference capability on unseen data.

 \begin{figure*}[htbp]
    \centering
    \includegraphics[width=0.9\textwidth]{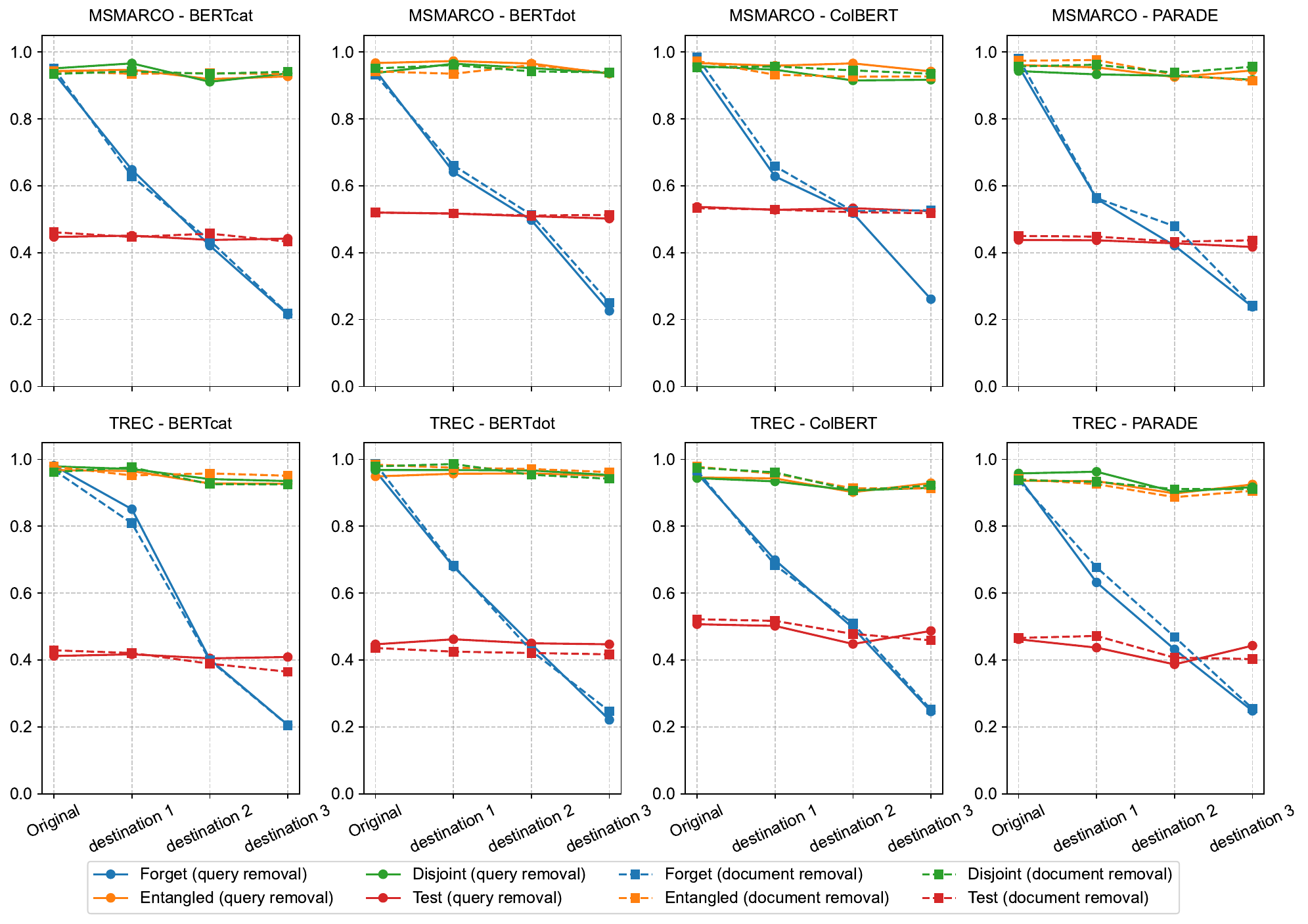}
    \caption{Performance shifts from $\trainedModel$ to the three forgetting destinations with a forget set size of 25\%.}
    \label{figure:controllable_forgetting}
\end{figure*}

\subsubsection{Controllable forgetting}

A key property of \gls{cocol} is its ability to achieve controllable forgetting. Specifically, by adjusting the target forget set performance specified in Eq.~\eqref{eq:mrr}, \gls{cocol} steers the performance of original model on the forget set toward the desired target, while maintaining stable performance on the retain and test sets. 

Beyond Goal~\ref{enum:goals:a}, we further align Eq.~\eqref{eq:mrr} with Goal~\ref{enum:goals:b} under two additional target values:

\begin{enumerate}[label=(\arabic*)]
    \item  the performance of $\retrainedModel$ on the forget set (typically higher than Goal~\ref{enum:goals:a}) to verify efficiency of \gls{cocol} in achieving similar performance to the $\retrainedModel$;
    \item half of the performance of the $\retrainedModel$ on the test set, to evaluate unlearning performance when requiring lower performance on the forget set.
\end{enumerate}

We sort these three target values by their theoretical MRR and refer to them as destination~1 (first item of Goal~\ref{enum:goals:b}), destination~2 (Goal~\ref{enum:goals:a}), and destination~3 (second item of Goal~\ref{enum:goals:b}). 

We conducted experiments on the 25\% forget set, with results summarised in Figure~\ref{figure:controllable_forgetting}. Although \gls{cocol} exhibits minor fluctuations in performance on the entangled, disjoint, and test sets across different destinations, its overall performance remains stable. This observation confirms \gls{cocol}’s ability to achieve controllable forgetting while preserving performance on retained and unseen data.

\subsubsection{Unlearning efficiency}

\begin{figure*}[htbp]
    \centering
    \includegraphics[width=0.9\textwidth]{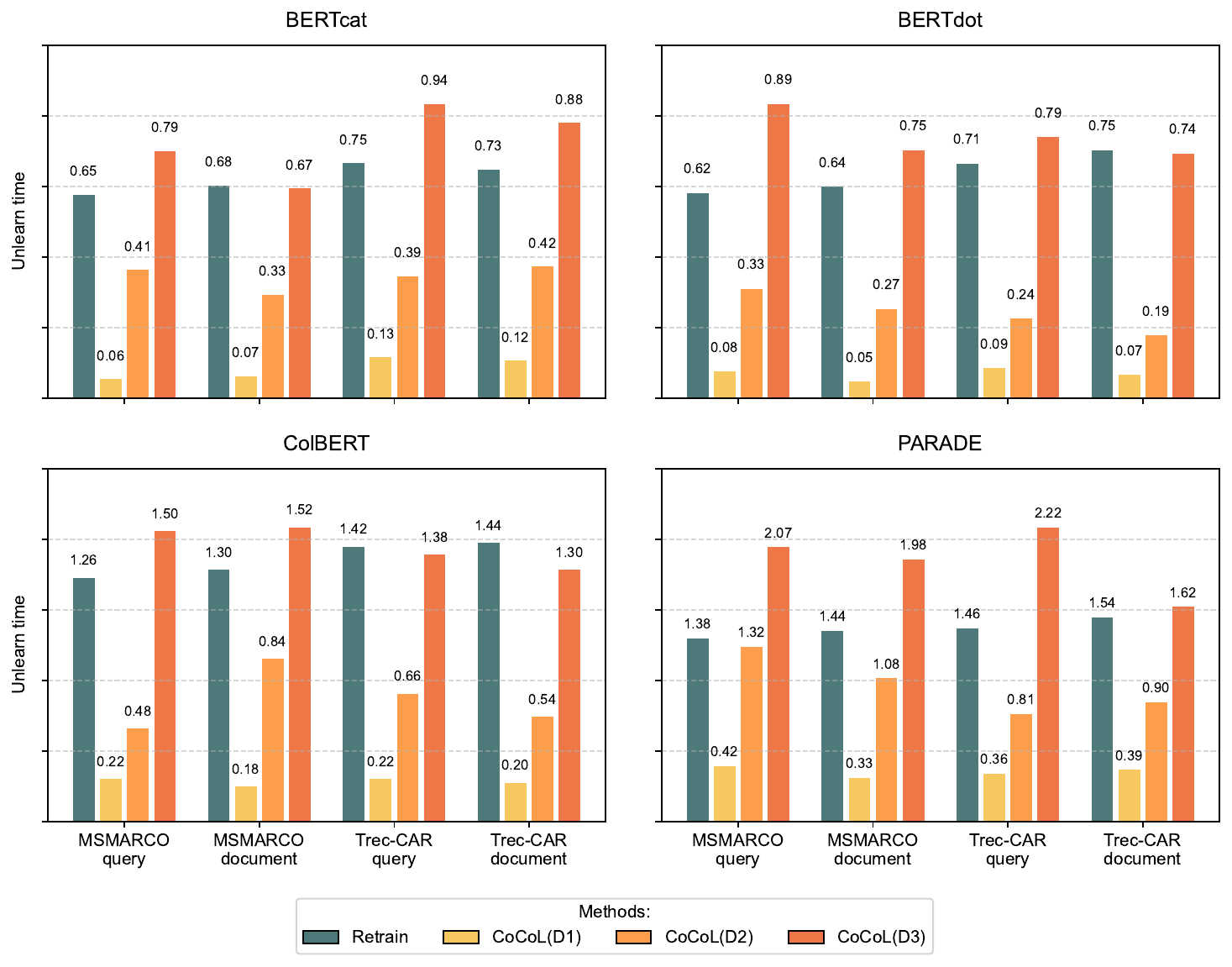}
    \caption{Comparison of unlearning time between Retrain and \gls{cocol} across different destinations. Here, CoCoL(D1), CoCoL(D2), and CoCoL(D3) indicate \gls{cocol}’s unlearning time for destinations~1, 2, and~3, respectively.}
    \label{figure:efficiency_comparison}
\end{figure*}

We further assessed the unlearning efficiency of \gls{cocol} relative to Retrain, with results summarised in Figure~\ref{figure:efficiency_comparison}. For destination~1, where \gls{cocol} is expected to match Retrain’s performance across all sets, it consistently achieves substantially lower unlearn time across all tasks and models. Similarly, for destination~2, which specifies a reduced target performance, \gls{cocol} still outperforms Retrain in terms of efficiency, underscoring its adaptability to different forgetting objectives. In contrast, for destination~3, where a substantially lower forget set performance is enforced, \gls{cocol} requires more runtime than Retrain on most tasks. 

\subsubsection{Ablation experiment}

This section describes the impact of individual loss components and parameter settings in \gls{cocol} on the unlearning performance.

To assess the effectiveness of each component in contrastive and consistent loss, we conducted an experiment with both query removal and document removal tasks from the \gls{marco} dataset using \gls{bertcat} neural-ranking model.

\begin{figure*}[htbp]
    \centering
    \includegraphics[width=0.9\textwidth]{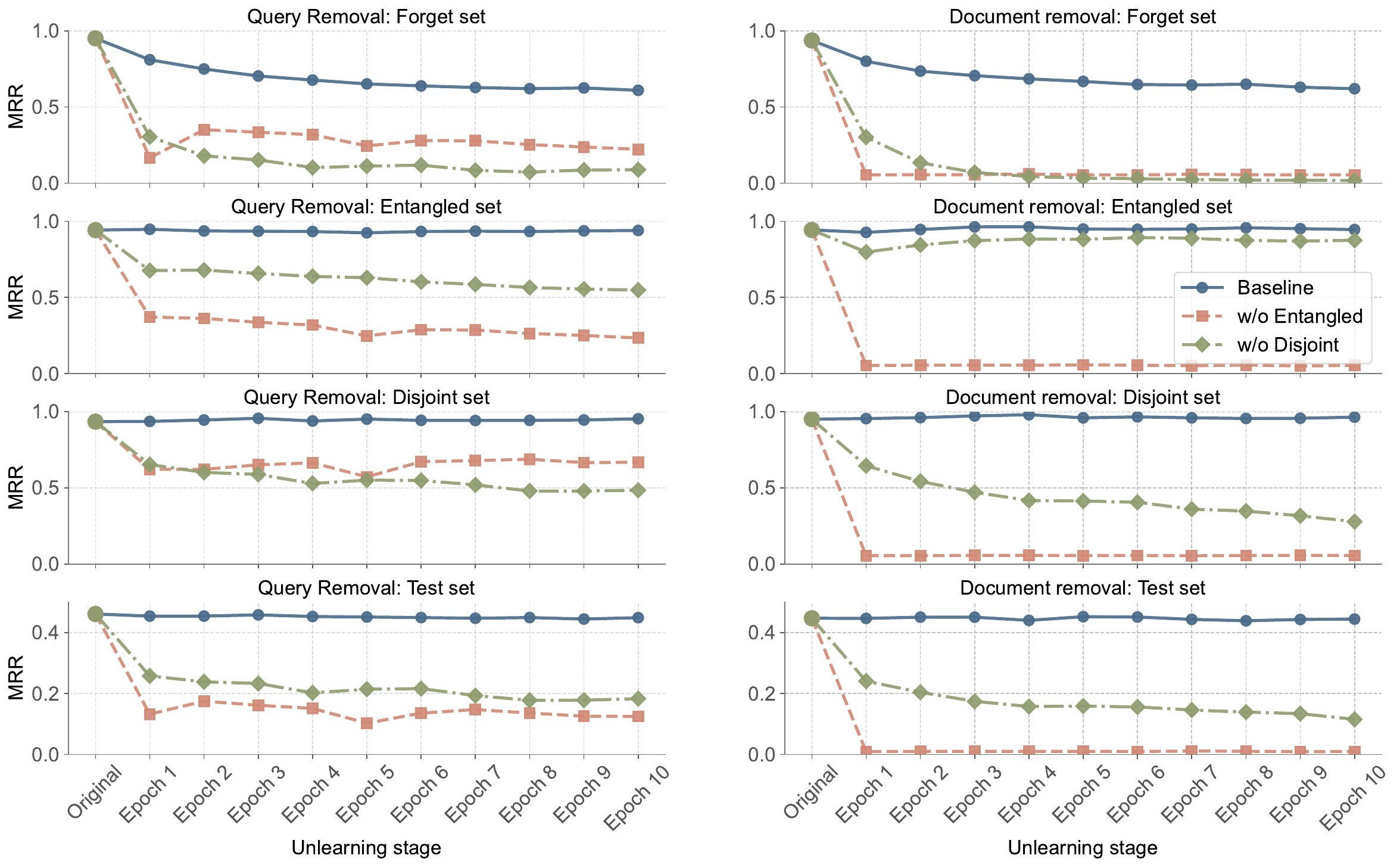}
    \caption{The impact of omitting different loss components. The `Baseline' indicates the benchmark performance using all loss components; `w/o Entangled' and `w/o Disjoint' denote the performance curves when excluding loss components associated with the entangled set in \eqref{eq:contrastive_loss} and the disjoint set in \eqref{eq:consistent_loss}, respectively. This figure demonstrates that both the entangled and consistent loss components are crucial for balancing forgetting performance with retention and inference capabilities.}
    \label{figure:ablation1}
\end{figure*}

Figure~\ref{figure:ablation1} illustrates that omitting the consistent loss led to a more rapid decline in \gls{mrr} scores on the forgetting set for both tasks, indicating that the consistent loss played a role in moderating the forgetting speed. The absence of consistent loss had a minimal impact on disjoint data: in query removal, the performance slightly underperformed the baseline model, whereas in document removal, it marginally surpassed the baseline.

The removal of the entangled component also resulted in an accelerated forgetting rate. However, this removal significantly diminished the performance of the unlearned model on all retained and unseen data, particularly in the query removal task, where there was a marked decline in model performance across all forget, retain, and test sets.

\subsection{Discussion}

Our experiments demonstrate that \gls{cocol} achieves superior unlearning performance across multiple forgetting objectives, maintaining alignment with $\retrainedModel$ on the forget set while preserving generalisation on retained and unseen data. By adjusting the target forget set performance, \gls{cocol} exhibits controllable forgetting, enabling fine-grained trade-offs between forgetting strength and model utility. Moreover, \gls{cocol} outperforms Retrain in efficiency, requiring substantially less runtime for most objectives. Notably, achieving stricter forgetting targets incurs higher computational costs, highlighting an inherent trade-off between forgetting precision and efficiency. These findings validate \gls{cocol}’s capacity as a flexible and scalable unlearning solution for neural \gls{ir} systems.

\subsubsection{Limitations}

Although \gls{cocol} achieves competitive unlearning performance across diverse neural IR models, it exhibits certain limitations. First, for traditional architectures such as Duet and MathPyramid, \gls{cocol} often requires substantially more training epochs to converge and struggles to maintain stable performance on the forget set, leading to inefficiencies and suboptimal trade-offs between forgetting and retention, as shown in Fig~\ref{fig:unlearning_process} in the Appendix. Second, while \gls{cocol} performs efficiently on PLM-based models under moderate forgetting targets, its runtime increases significantly when enforcing very low performance on the forget set, indicating a need for improved optimisation strategies under stricter constraints.

\subsubsection{Future work}
The exploration of \gls{cocol} capabilities has opened avenues for future research in \gls{nmu}. One critical area of focus should aim to distinguish between query removal and document removal more precisely. Recognizing and addressing the subtle differences between these two sub-tasks could lead to the development of more nuanced and targeted unranking methods, enhancing the overall effectiveness and accuracy of the \gls{nmu} task.

In addition, considering that neural ranking models typically comprise both embedding and ranking modules, it is imperative to investigate how unranking methods interact with these components differently. Future research should delve into the distinct impacts of unranking on embedding and ranking modules, and accordingly, develop improved unranking methods that treat these modules differently. Such an approach could lead to more effective unranking techniques, further advancing \gls{nmu}.

\glsreset{ir}
\glsreset{nmu}
\glsreset{cocol}

\section{Conclusion} \label{sec:conclu}

In an era where data privacy and dynamic information landscapes are paramount, this study focuses on the field of machine unlearning, specifically within the context of neural ranking models for \gls{ir} systems. This research introduced the concept of \gls{nmu}, presenting a novel method (\gls{cocol}) that effectively balances the delicate trade-off between controllable forgetting specific information and maintaining the overall performance of neural ranking models. \Gls{cocol} is  effective with pretraining-based neural ranking models, representing an advancement in addressing the unique challenges posed by machine unlearning in \gls{ir} systems.

\section*{Acknowledgement}
This work was supported by the China Postdoctoral Science Foundation under Grant No. 2025M773205, the Postdoctoral Fellowship Program of CPSF under Grant No. GZC20252319, and the National Natural Science Foundation of China under Grant No. 72374161. The authors would also like to thank the editors and anonymous reviewers for their valuable comments and suggestions, which greatly improved the quality of this manuscript.

\section*{Data availability}
 To access the dataset and reproduce the experiments, please refer to the paper's GitHub repository located at
\href{https://github.com/JingruiHou/NuMuR}{github.com/JingruiHou/NuMuR}.

\appendix

\section*{Relevance score interval comparison} \label{sec:re.comparison}

\begin{figure}[htbp]
\centering
        \includegraphics[width=0.45\textwidth]{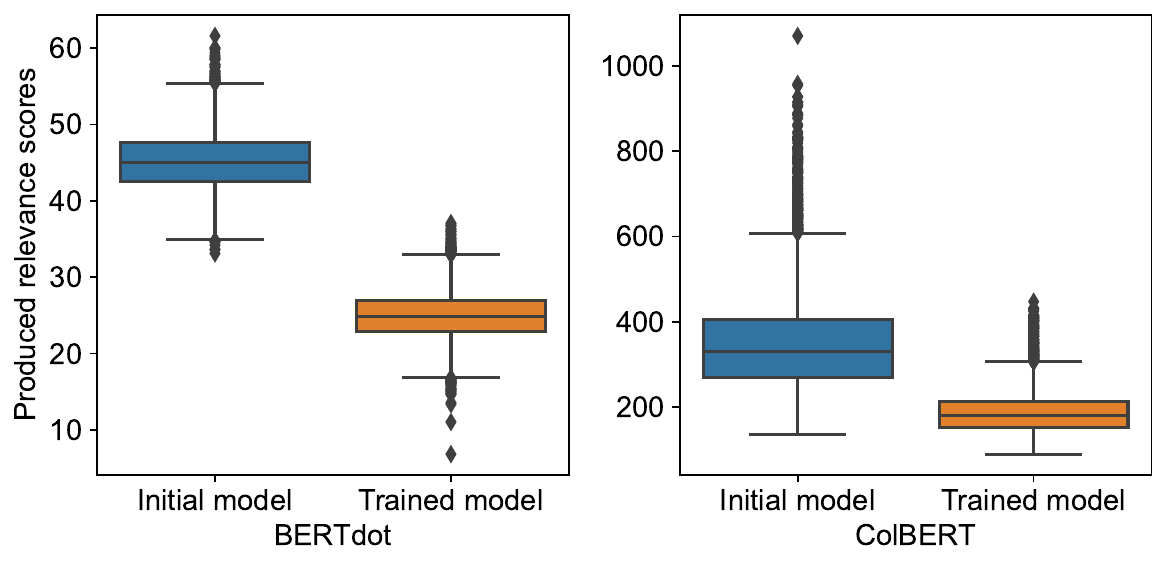} 
    \caption{Comparison between $\initialModel$ and $\trainedModel$.  The figure shows that neural ranking models trained on certain datasets may experience shifts in relevance score intervals. }
    \label{figure:compare_train_init}
\label{figure:rel_comparison}
\end{figure}

This section provides examples showing that relevance score distributions vary across different neural ranking models. Figure~\ref{figure:compare_train_init} illustrates differences in the scale of relevance scores. Both \gls{bertdot} and \gls{colbert} exhibit relatively lower relevance score ranges after training. Using $\initialModel$ as the `incompetent' teacher may result in higher relevance scores on forgetting samples. Additionally.

\section*{Detailed experimental result}

Table~\ref{tab:detail_maro} and Table~\ref{tab:detail_trec} provide comprehensive visualisations of unlearning performance for \gls{cocol} and baseline methods across two datasets (\gls{marco} and \gls{trec}), three forget set sizes, and four neural \gls{ir} models. Detailed performance scores for all experiments are listed in Table~\ref{tab:msmarco_results} and Table~\ref{tab:trec_results}.

\begin{table*}[htbp]
    \centering
\caption{Unlearning performance across models for varying forget set sizes on \gls{marco}. \#QR and \#DR denote query removal and document removal, respectively. Forget set performance is normalised as $1 - |\text{MRR}^{\unlearnedModel}_{\text{forget}} - \text{MRR}^{\retrainedModel}_{\text{test}}|$, where $\text{MRR}^{\retrainedModel}_{\text{test}}$ is the MRR of $\retrainedModel$ on the test set. Higher values indicate closer alignment with $\retrainedModel$ (ideal performance on the vertex). The Y-axis ranges from 0 to 1 for all sets except the test set (0 to 0.5).}

    \renewcommand{\arraystretch}{1.2}
    \begin{tabular}{|c|c|c|c|}
        \hline
        \textbf{Model} & \textbf{5\%} & \textbf{15\%} & \textbf{25\%} \\
        \hline
        BERTcat
        & \includegraphics[width=0.25\textwidth]{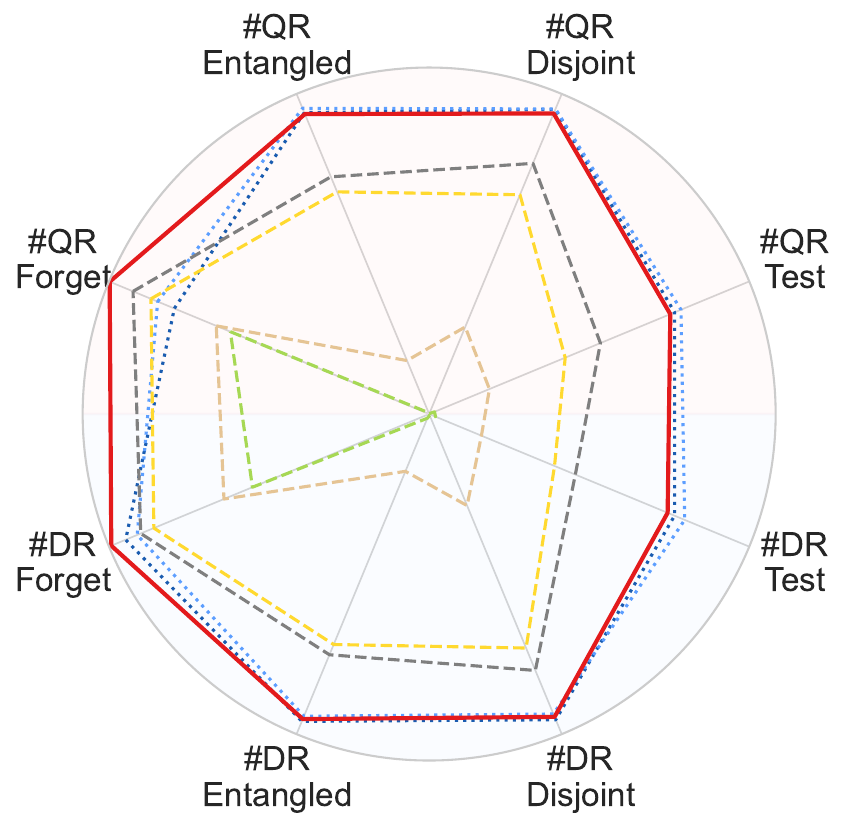}
        & \includegraphics[width=0.25\textwidth]{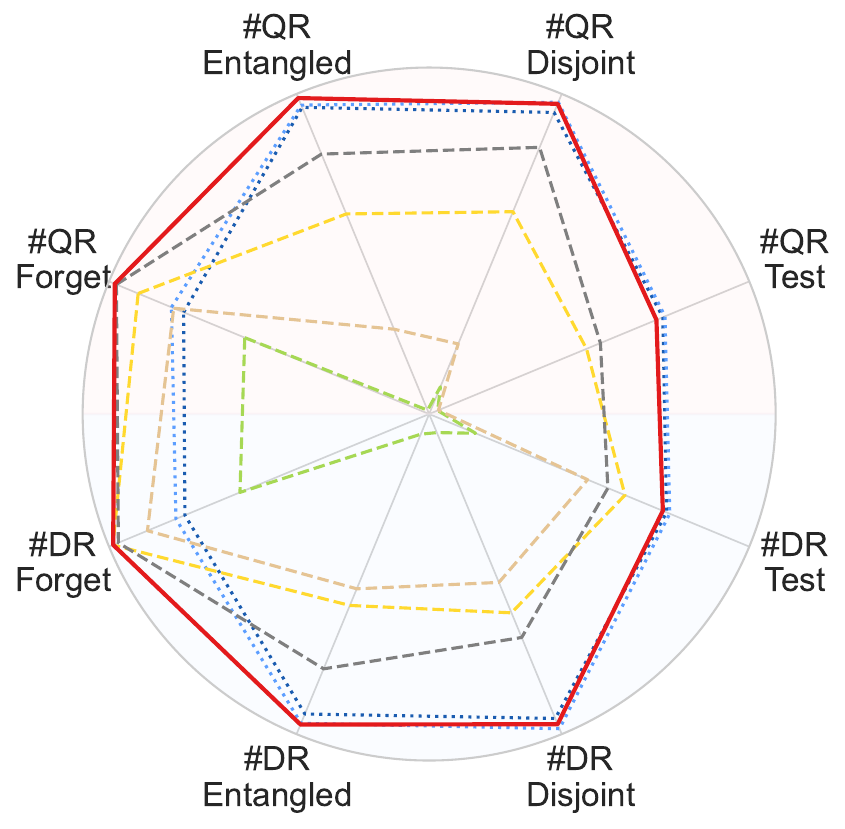}
        & \includegraphics[width=0.25\textwidth]{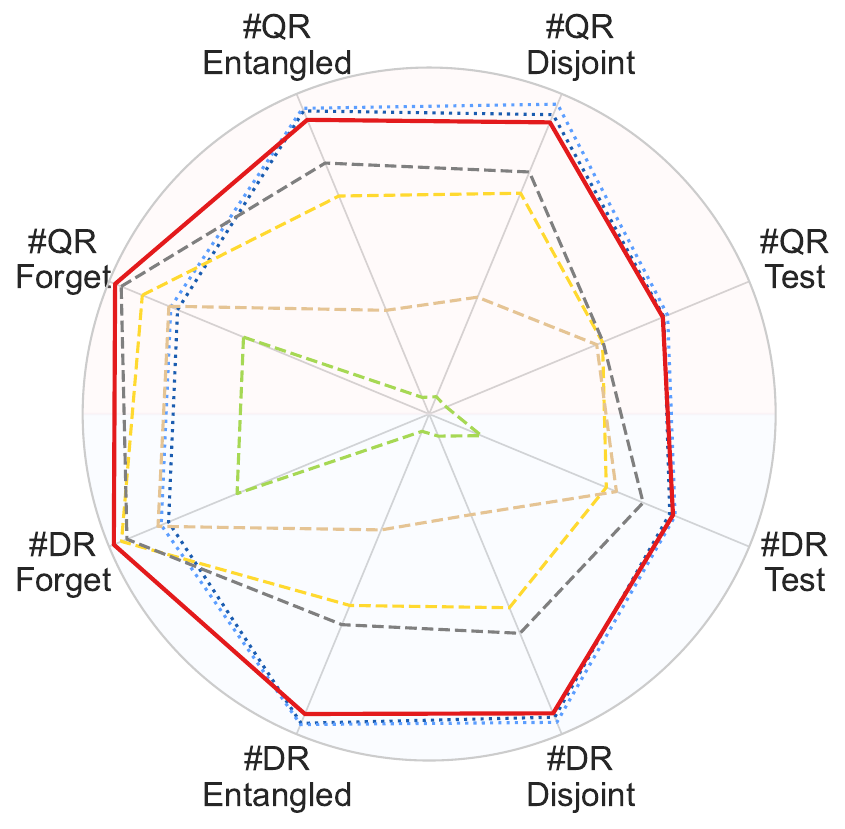} \\
        \hline
        BERTdot
        & \includegraphics[width=0.25\textwidth]{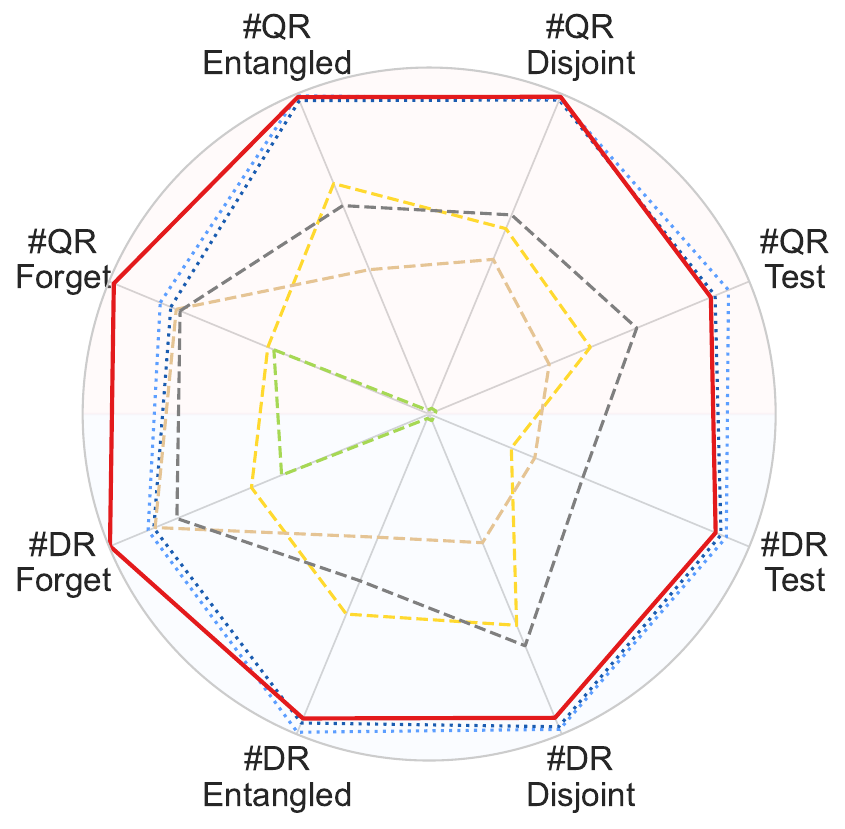}
        & \includegraphics[width=0.25\textwidth]{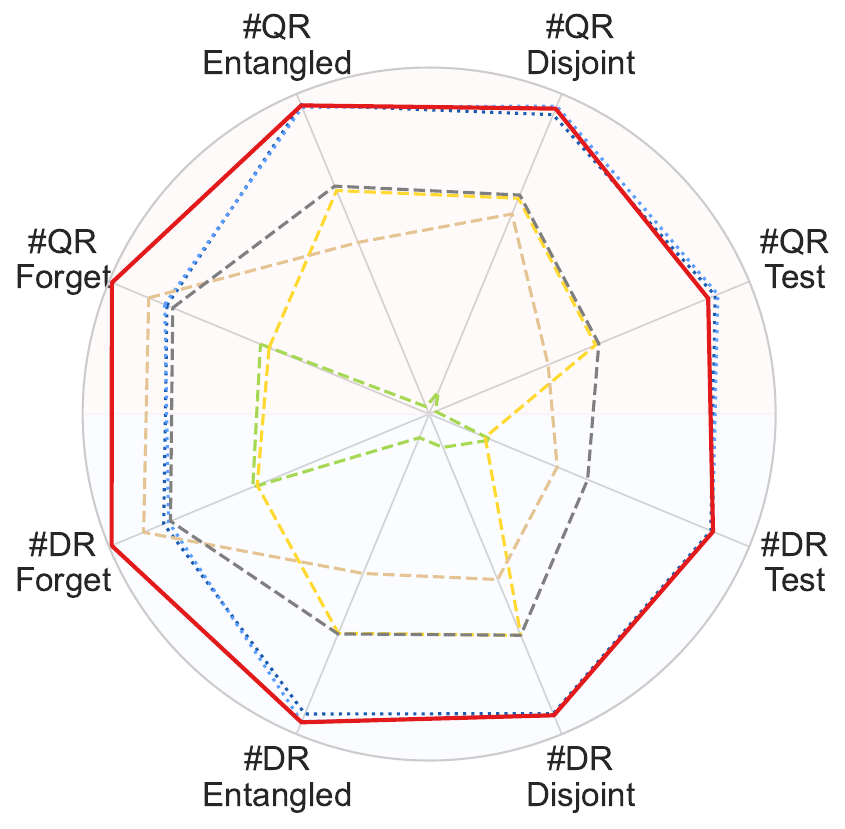}
        & \includegraphics[width=0.25\textwidth]{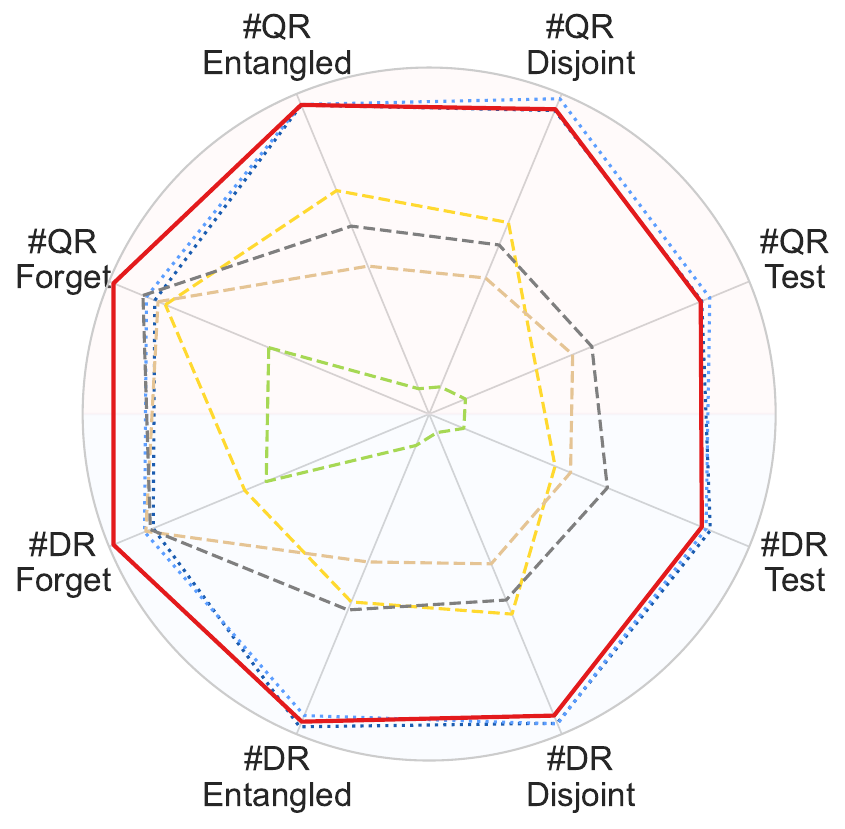} \\
        \hline
        ColBERT
        & \includegraphics[width=0.25\textwidth]{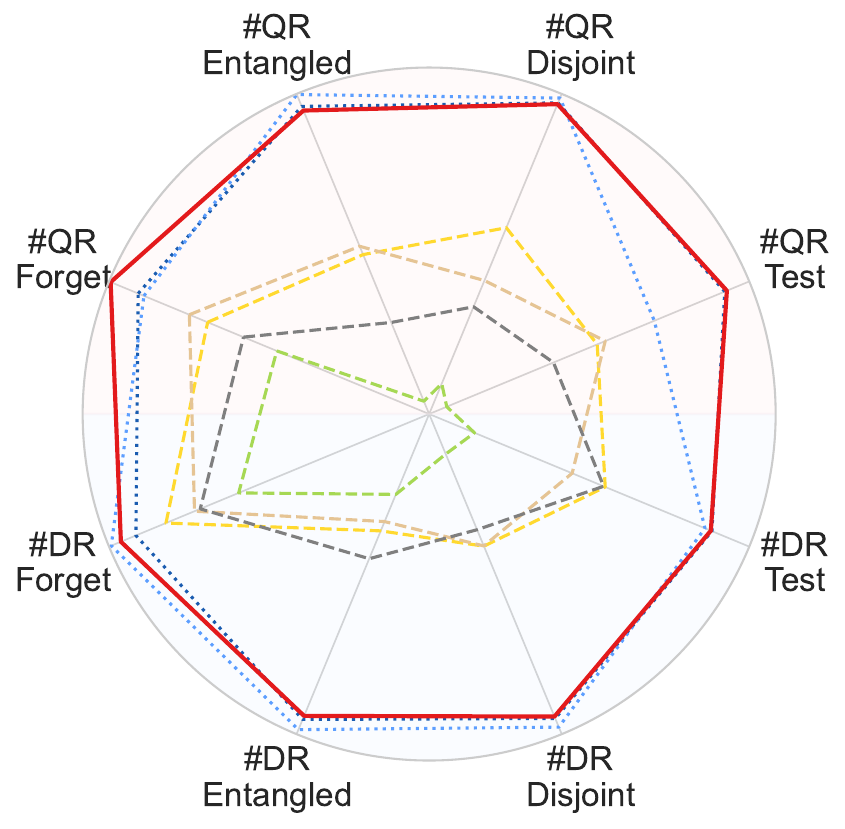}
        & \includegraphics[width=0.25\textwidth]{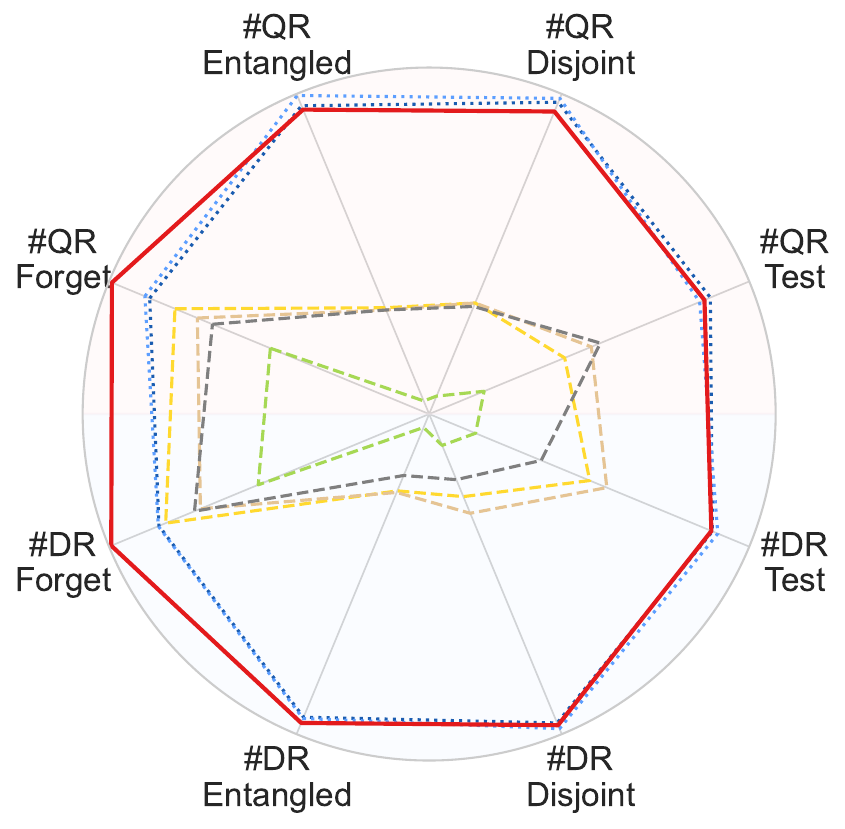}
        & \includegraphics[width=0.25\textwidth]{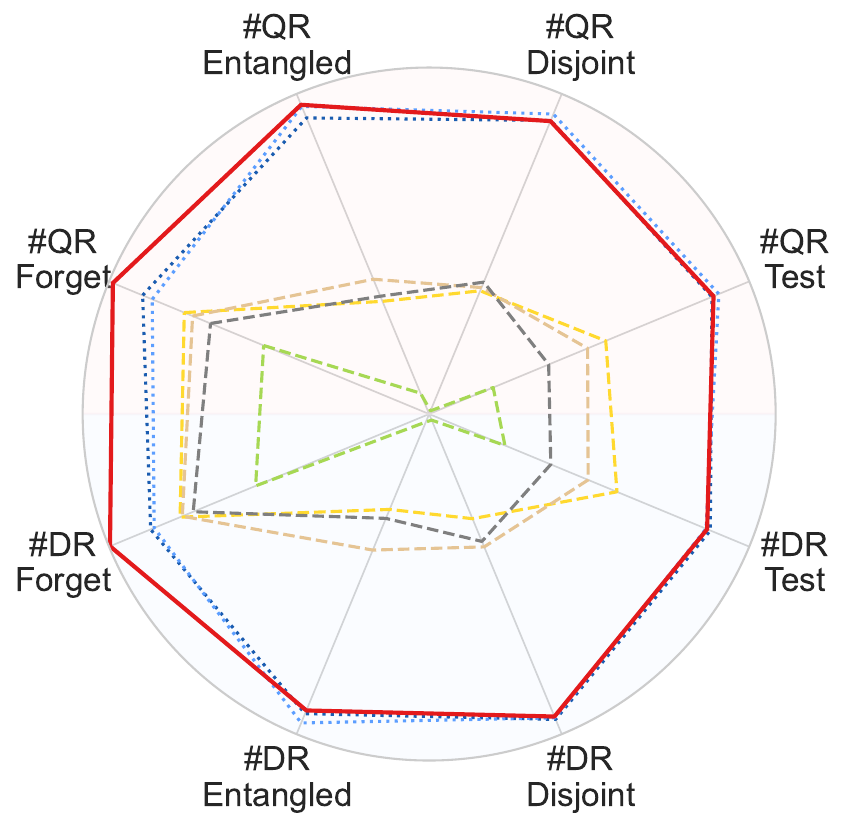} \\
        \hline
        PARADE
        & \includegraphics[width=0.25\textwidth]{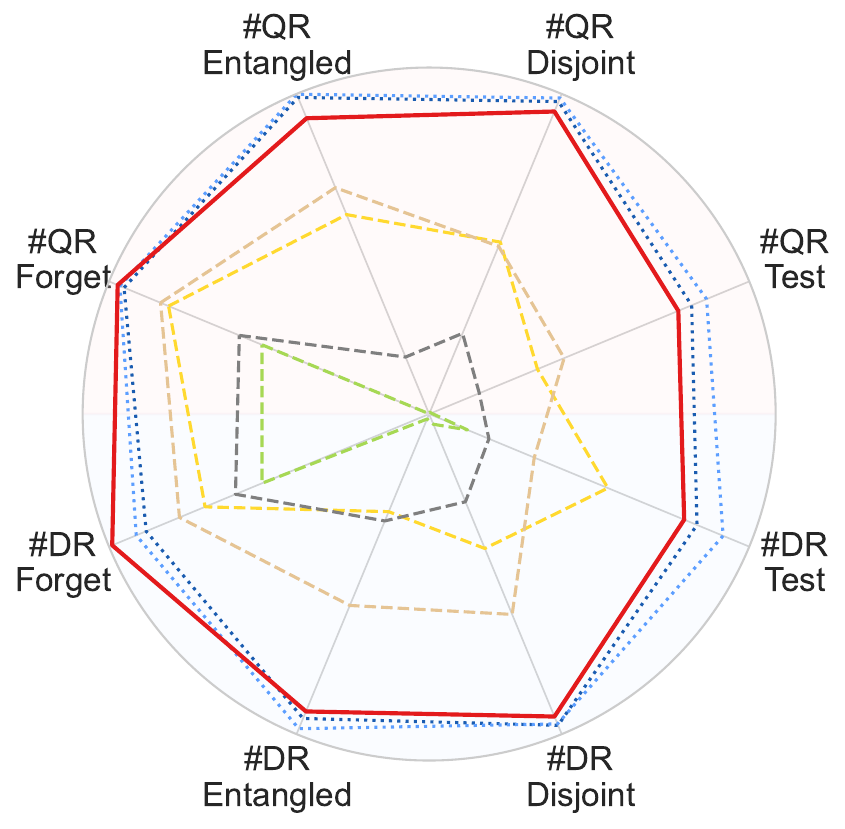}
        & \includegraphics[width=0.25\textwidth]{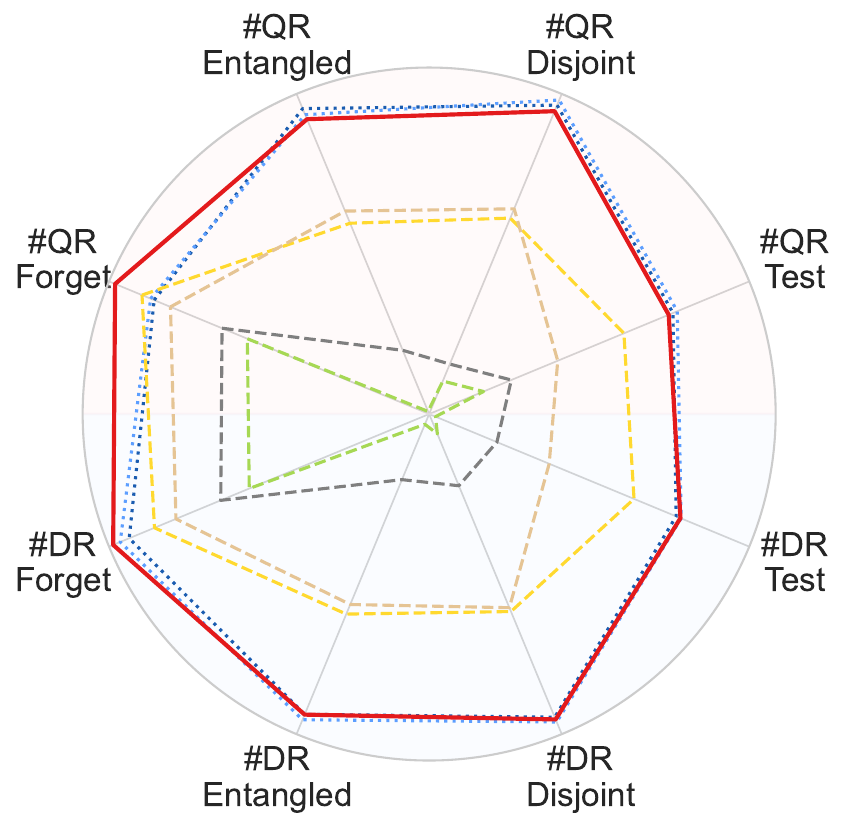}
        & \includegraphics[width=0.25\textwidth]{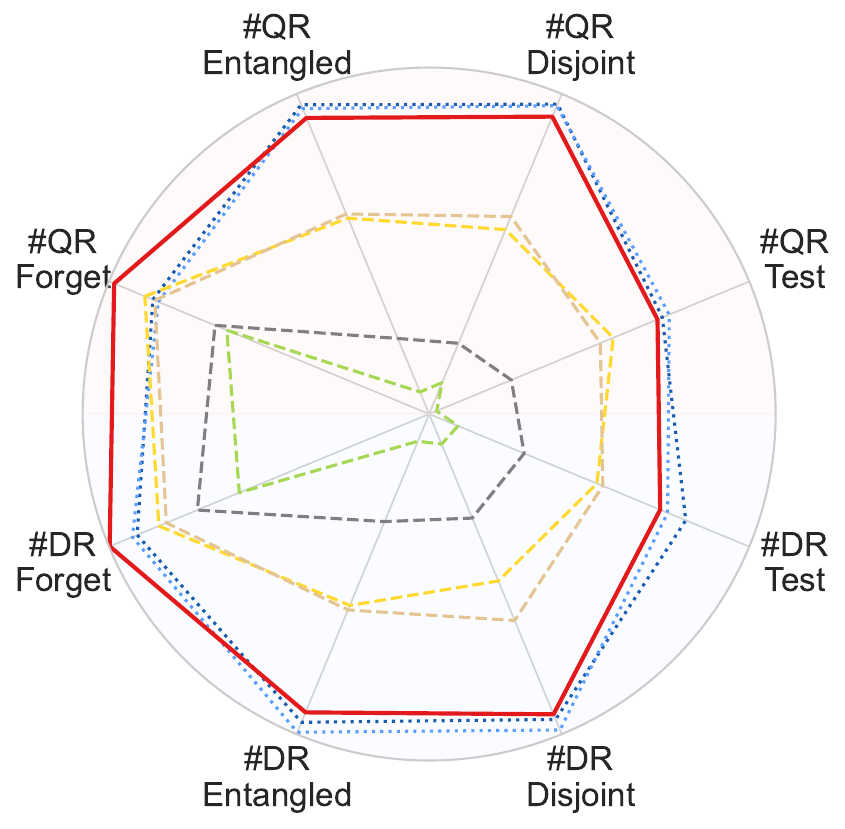} \\
        \hline
        \multicolumn{4}{|c|}{\includegraphics[width=0.8\textwidth]{figs/radar_plots_sci/legend_horizontal.pdf}} \\
        \hline
    \end{tabular}
    \label{tab:detail_maro}
\end{table*}

\begin{table*}[htbp]
    \centering
    \caption{Unlearning performance across models for varying forget set sizes on \gls{trec}. The radar charts use the same notation as described in the caption of Table~\ref{tab:detail_maro}.}

    \renewcommand{\arraystretch}{1.2}
    \begin{tabular}{|c|c|c|c|}
        \hline
        \textbf{Model} & \textbf{5\%} & \textbf{15\%} & \textbf{25\%} \\
        \hline
        BERTcat
        & \includegraphics[width=0.25\textwidth]{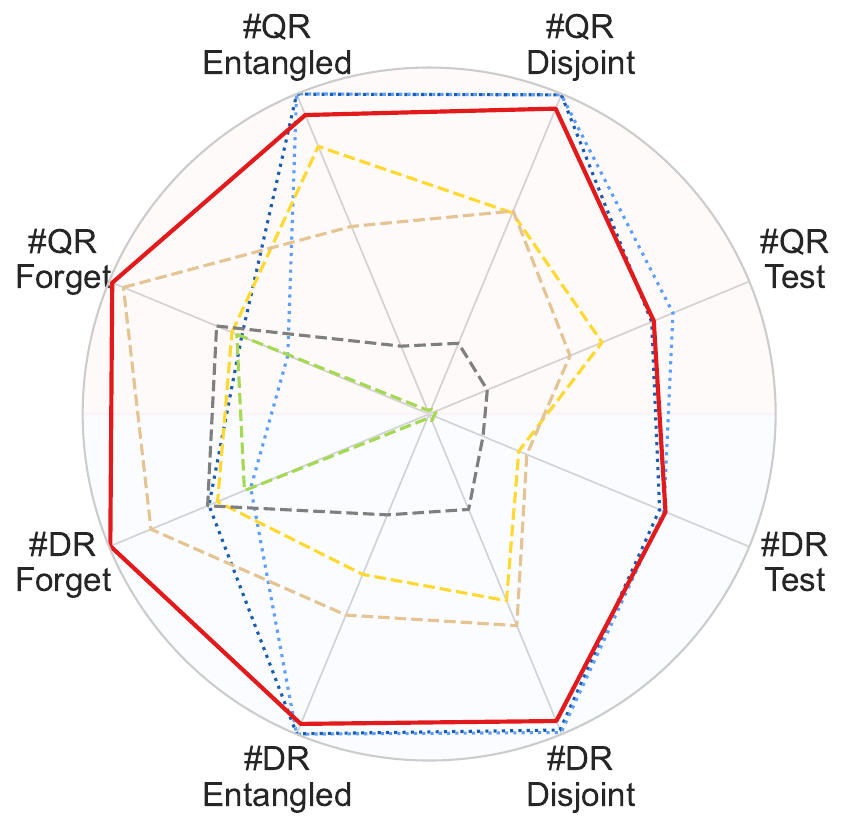}
        & \includegraphics[width=0.25\textwidth]{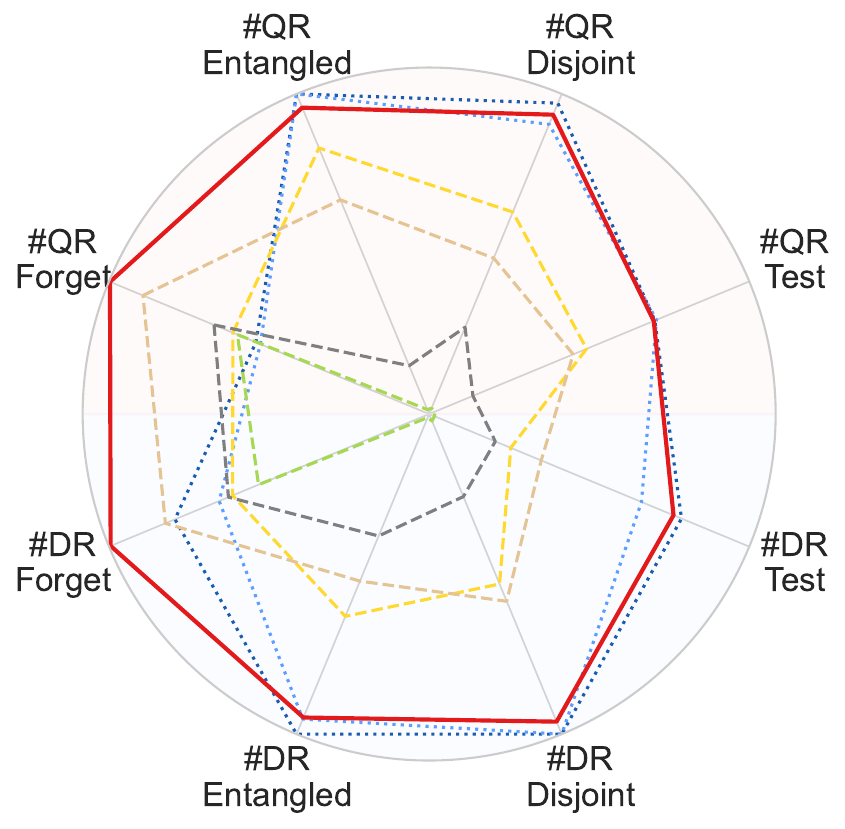}
        & \includegraphics[width=0.25\textwidth]{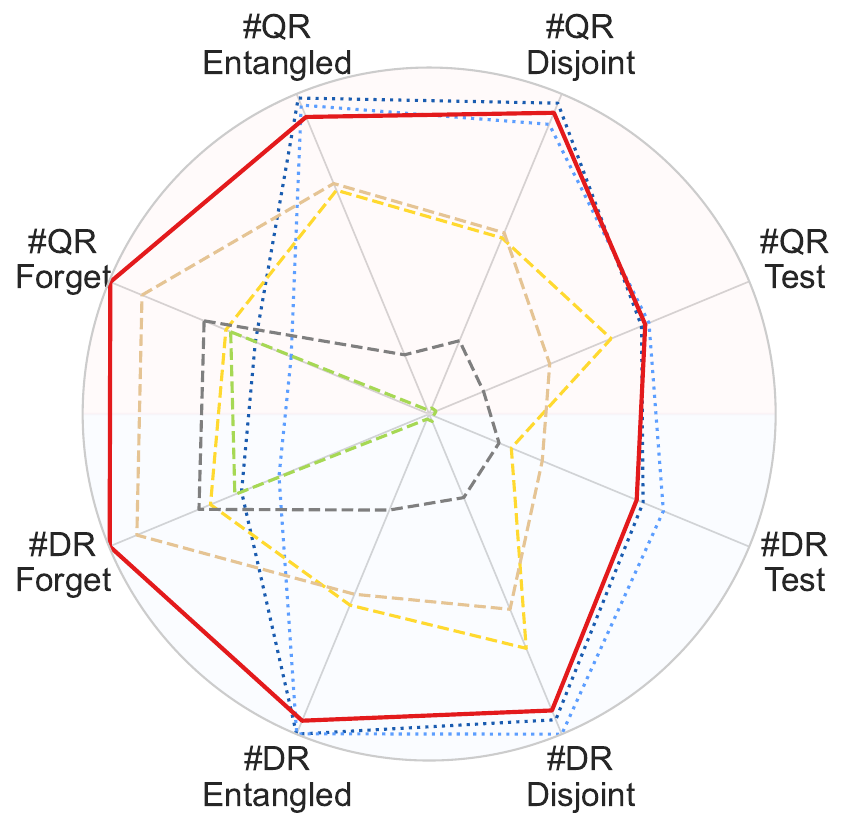} \\
        \hline
        BERTdot
        & \includegraphics[width=0.25\textwidth]{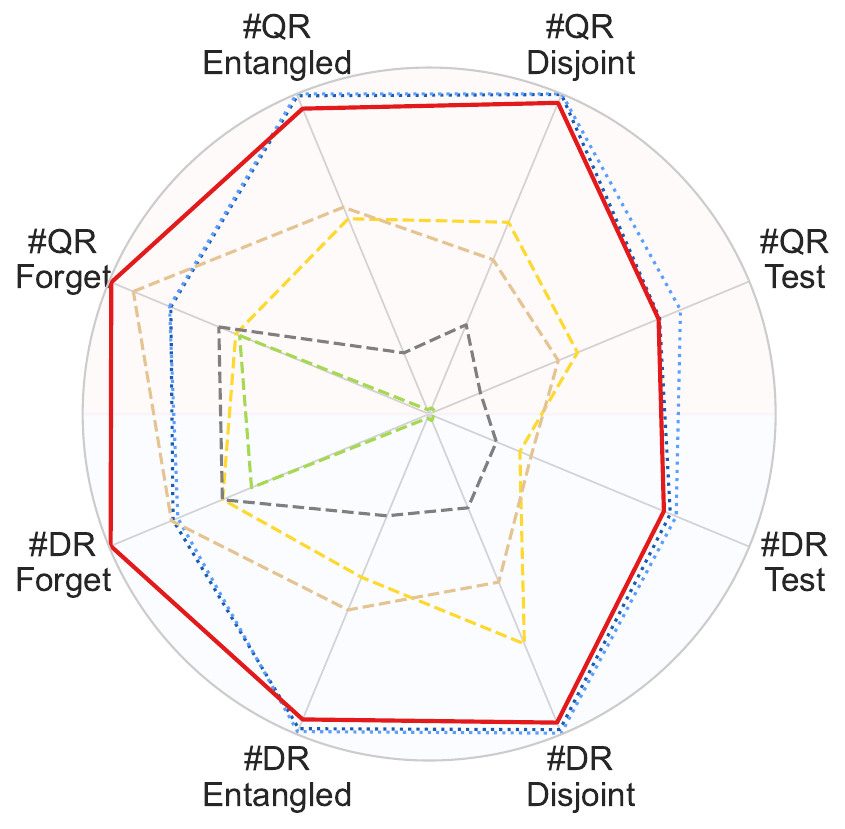}
        & \includegraphics[width=0.25\textwidth]{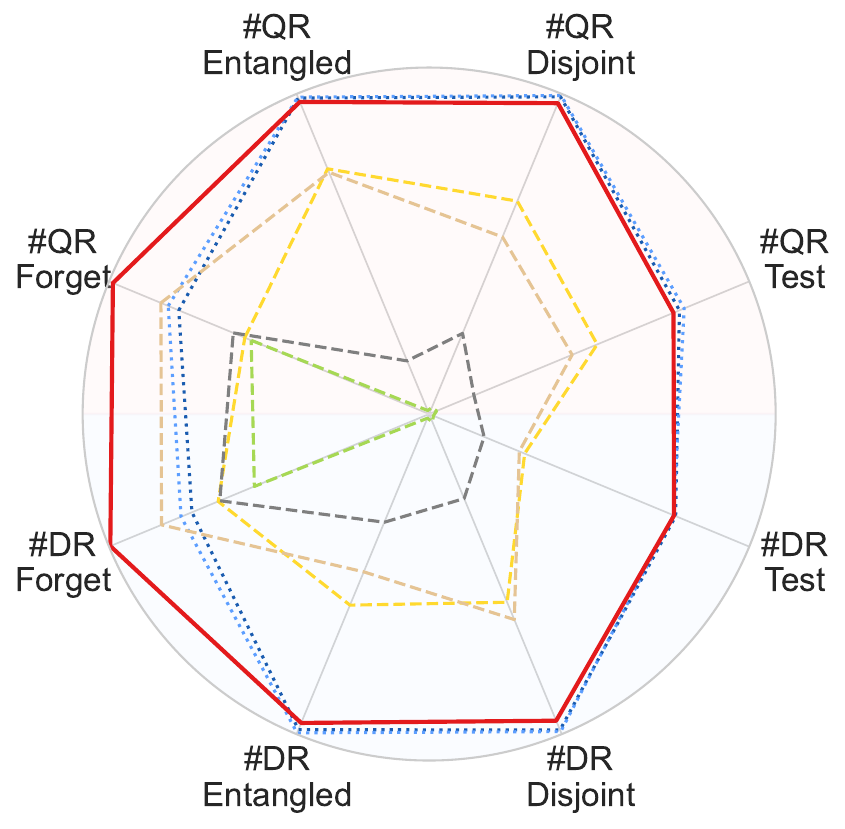}
        & \includegraphics[width=0.25\textwidth]{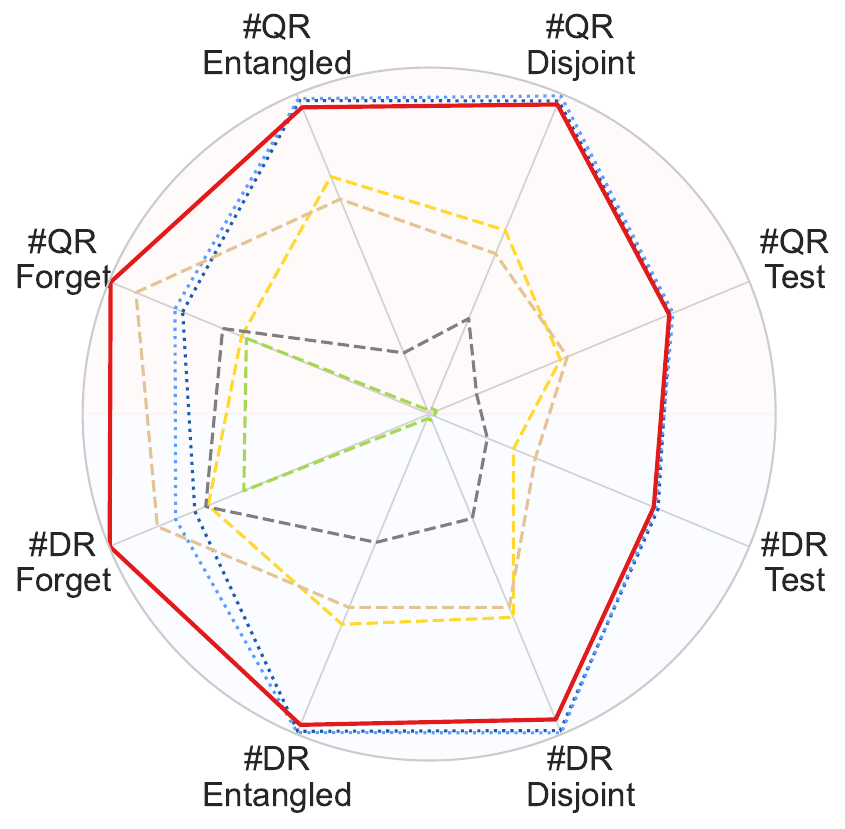} \\
        \hline
        ColBERT
        & \includegraphics[width=0.25\textwidth]{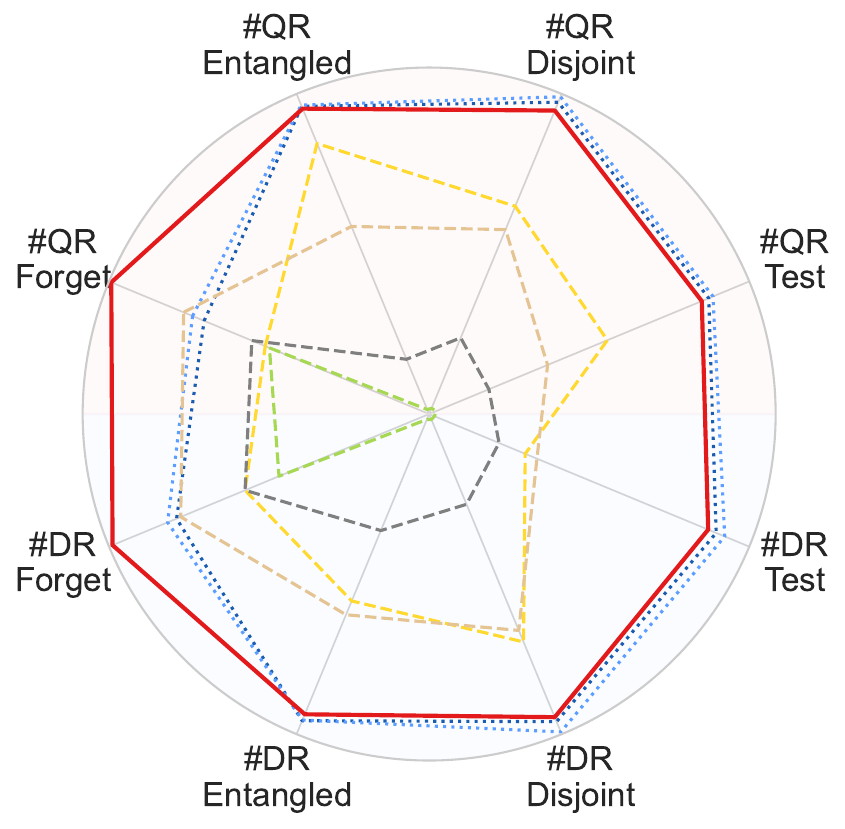}
        & \includegraphics[width=0.25\textwidth]{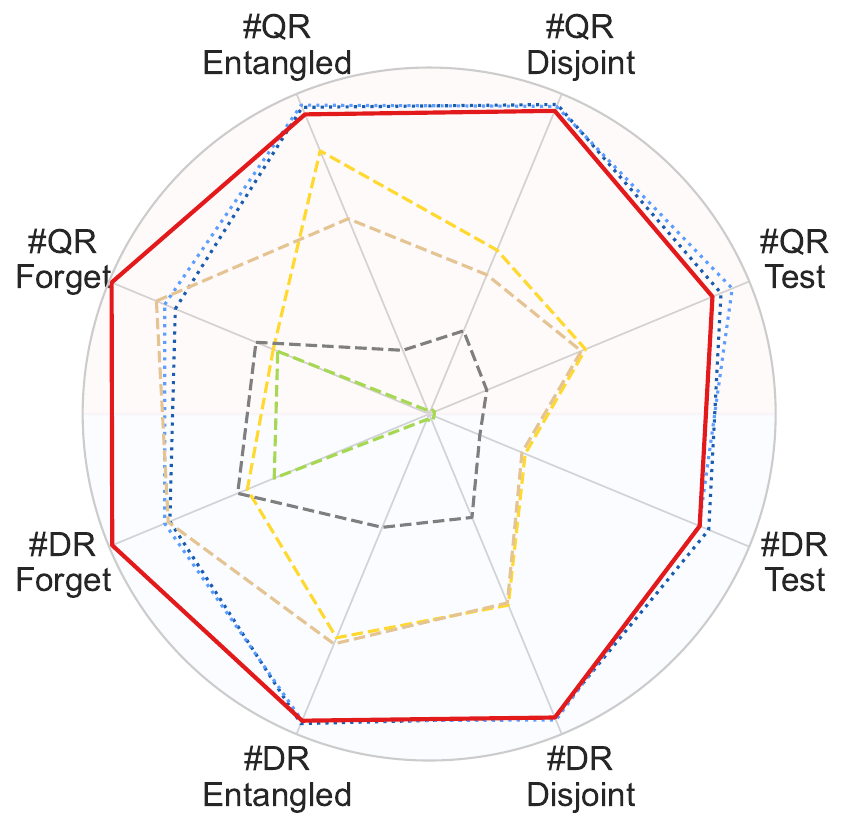}
        & \includegraphics[width=0.25\textwidth]{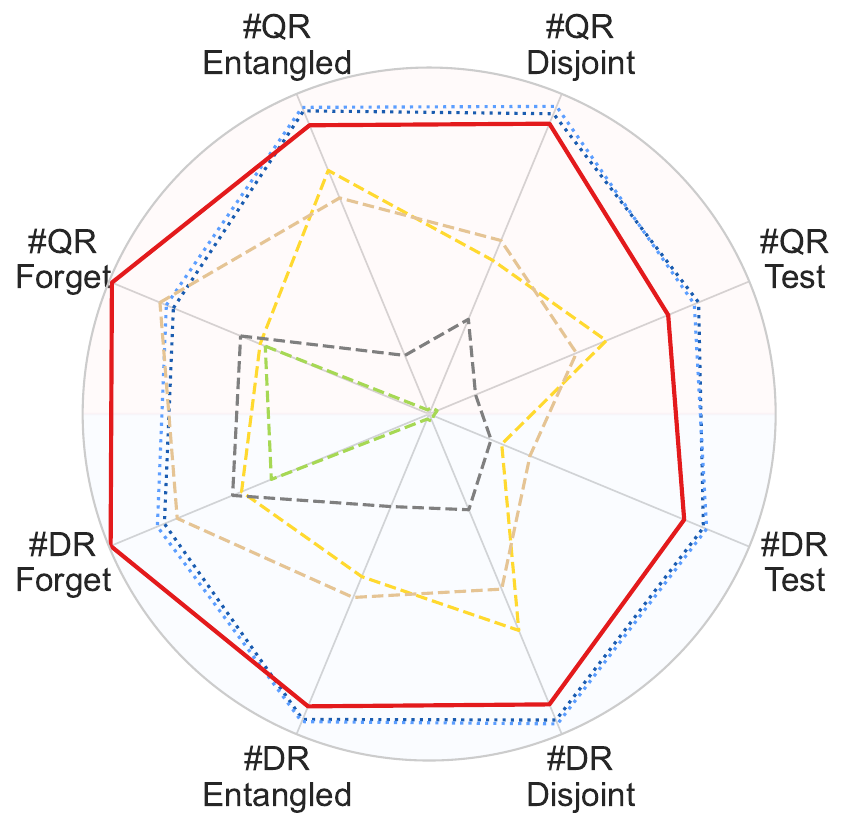} \\
        \hline
        PARADE
        & \includegraphics[width=0.25\textwidth]{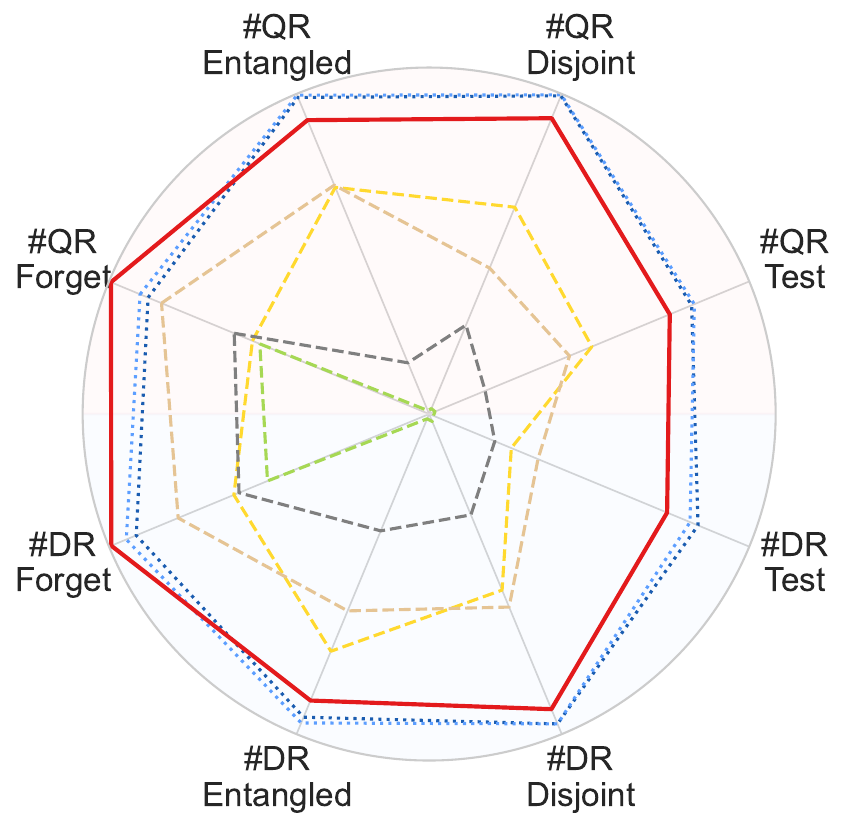}
        & \includegraphics[width=0.25\textwidth]{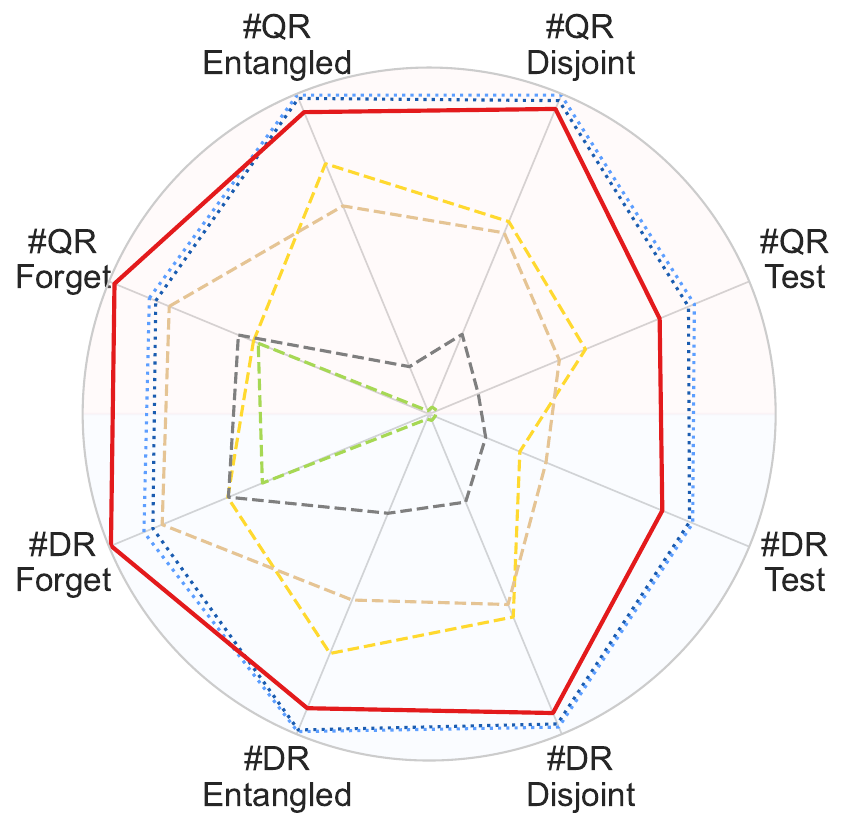}
        & \includegraphics[width=0.25\textwidth]{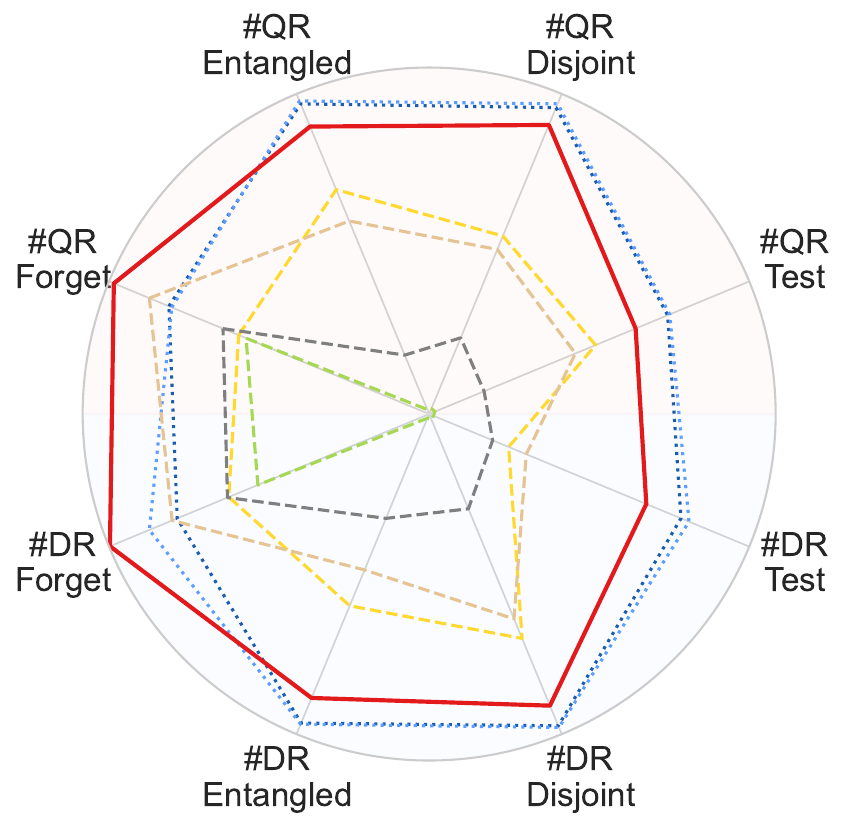} \\
        \hline
    \end{tabular}
    \label{tab:detail_trec}
\end{table*}

\section*{Additional experiment}

\begin{figure}[htbp]
    \begin{subfigure}{.22\textwidth}
        \centering
        \includegraphics[width=1\textwidth]{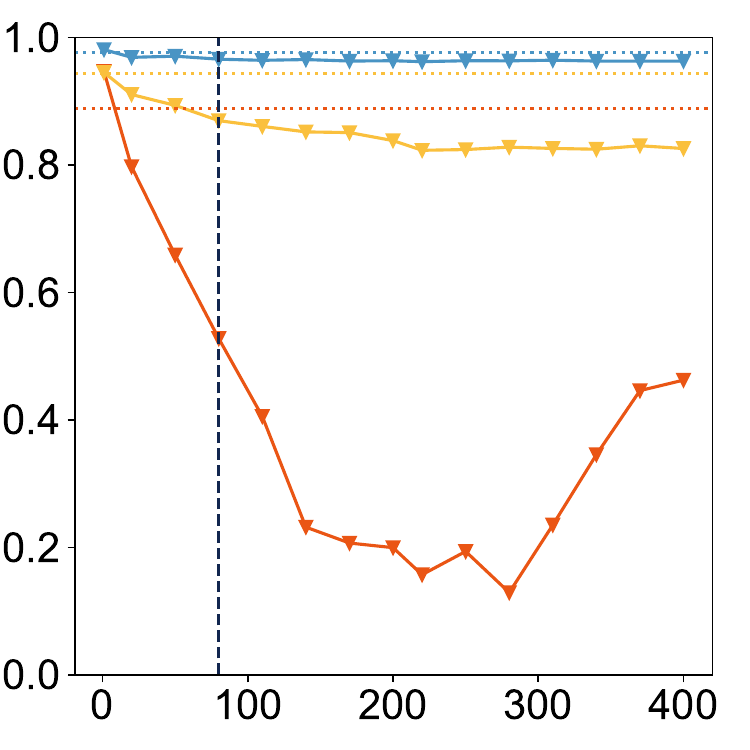} 
    \caption{\gls{trec} query removal with \gls{duet}}
    \label{figure.4(01)}
    \end{subfigure}
    \hfil
    \begin{subfigure}{.22\textwidth}
        \centering
        \includegraphics[width=1\textwidth]{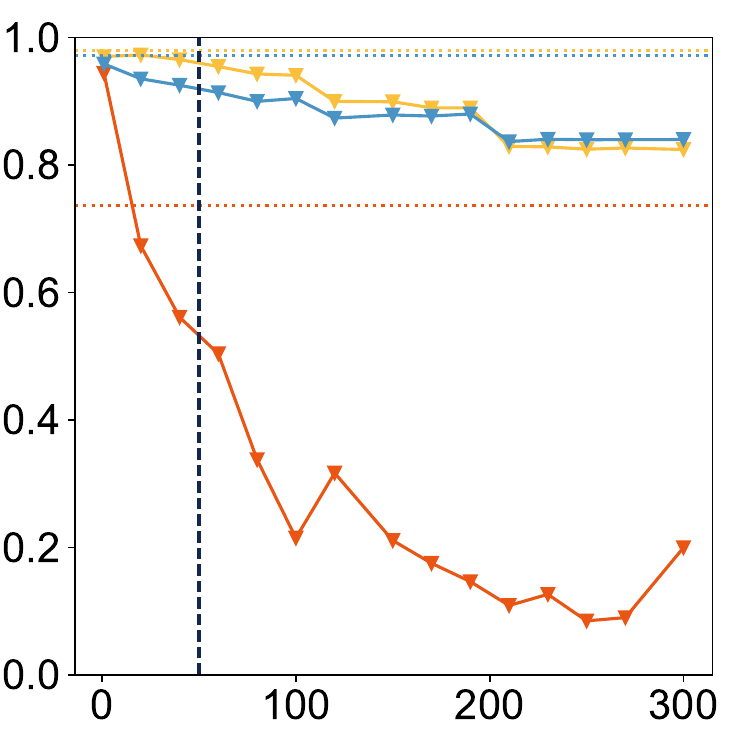}
        \caption{\gls{trec} document removal with \gls{duet}}
        \label{figure.4(02)}
    \end{subfigure}
    \begin{subfigure}{.22\textwidth}
        \centering
        \includegraphics[width=1\textwidth]{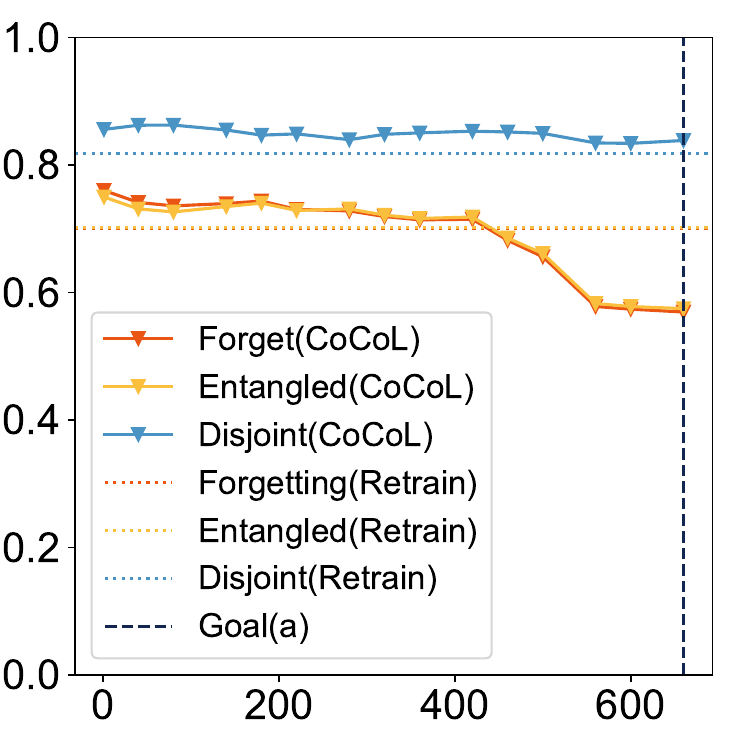}
    \caption{\gls{trec} query removal with \gls{matchp}}
    \label{figure.4(03)}
    \end{subfigure}
    \hfil
    \begin{subfigure}{.22\textwidth}
        \centering
        \includegraphics[width=1\textwidth]{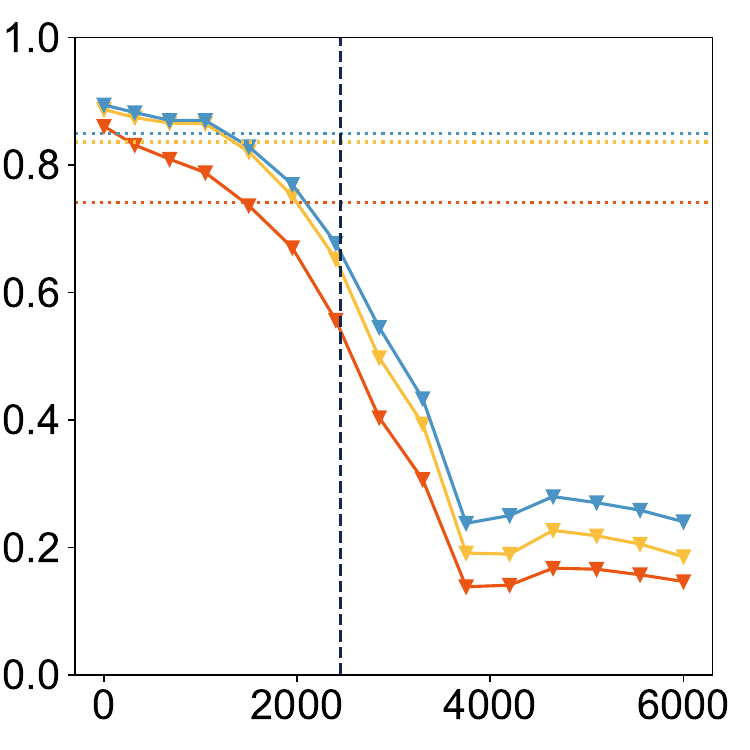}
        \caption{\gls{trec} document removal with \gls{matchp}}
        \label{figure.4(04)}
    \end{subfigure}
   
\caption{Unlearning performance of \gls{cocol} across various traditional neural ranking models on different tasks.} 
\label{fig:unlearning_process}
\end{figure}

Figure~\ref{fig:unlearning_process} illustrates the unlearning performance of two traditonal neural ranking models (\gls{duet} and \gls{matchp}) across different tasks. The `Ideal epochs' correspond to the stopping points that satisfy Goal~\ref{enum:goals:a}. 

While \gls{cocol} demonstrates strong overall performance, it exhibits limitations on certain traditional models. For instance, when unlearning a trained \gls{duet} model, \gls{cocol} requires hundreds of epochs to converge, and results in inefficiency and fluctuating performance on the forget set. Similarly, for \gls{matchp}, it struggles to achieve a precise balance between forgetting and retaining, leading to suboptimal trade-offs in model utility.

\begin{sidewaystable}[htbp]
\caption{
\textbf{Detailed Performance of Unlearning Methods on \gls{marco} Dataset.} 
Results are reported across varying removal ratios (5\%, 15\%, 25\%) and removal types (query removal vs.\ document removal). 
Evaluation is conducted on four disjoint subsets: 
\textbf{F} (Forget set), \textbf{E} (Entangled set), \textbf{D} (Disjoint set), and \textbf{T} (Test set).}
\resizebox{\textheight}{!}{%
\begin{tabular}{@{}|l|llllllll|l|llllllll|l|llllllll|@{}}
\toprule
\multicolumn{1}{c}{} & \multicolumn{4}{c}{\textbf{Query removal}}  & \multicolumn{4}{c}{Document removal} & \multicolumn{1}{c}{} & \multicolumn{4}{c}{\textbf{Query   removal}}  & \multicolumn{4}{c}{\textbf{Document removal}} & \multicolumn{1}{c}{} & \multicolumn{4}{c}{\textbf{Query removal}}  & \multicolumn{4}{c}{\textbf{Document removal}} \\
                     & F & E & D & T & F  & E  & D & T &                      & F & E & D & T & F  & E  & D & T &                      & F & E & D & T & F & E  & D & T \\ \cmidrule(lr){2-9} \cmidrule(lr){11-18} \cmidrule(lr){19-26}

\multicolumn{9}{c}{BERTcat 5\%}                                                                      & \multicolumn{9}{c}{BERTcat 15\%}                                                                     & \multicolumn{9}{c}{BERTcat 25\%}                                                                     \\
Retrain              & 0.66   & 0.94      & 0.95     & 0.46 & 0.51    & 0.96       & 0.96     & 0.46 &                      & 0.67   & 0.96      & 0.94     & 0.44 & 0.68    & 0.94       & 0.95     & 0.45 &                      & 0.65   & 0.95      & 0.94     & 0.44 & 0.64    & 0.97       & 0.95     & 0.45 \\
CF                   & 0.61   & 0.96      & 0.95     & 0.47 & 0.55    & 0.94       & 0.94     & 0.48 &                      & 0.63   & 0.97      & 0.97     & 0.44 & 0.65    & 0.97       & 0.98     & 0.45 &                      & 0.63   & 0.96      & 0.97     & 0.45 & 0.61    & 0.97       & 0.96     & 0.46 \\
Amnesiac             & 0.08   & 0.00      & 0.00     & 0.01 & 0.01    & 0.01       & 0.00     & 0.01 &                      & 0.01   & 0.01      & 0.09     & 0.02 & 0.04    & 0.06       & 0.06     & 0.09 &                      & 0.02   & 0.05      & 0.05     & 0.03 & 0.05    & 0.05       & 0.07     & 0.10 \\
NegGrad              & 0.33   & 0.69      & 0.69     & 0.26 & 0.32    & 0.72       & 0.73     & 0.24 &                      & 0.53   & 0.63      & 0.63     & 0.30 & 0.46    & 0.60       & 0.62     & 0.37 &                      & 0.34   & 0.68      & 0.69     & 0.33 & 0.41    & 0.60       & 0.61     & 0.33 \\
SSD                  & 0.13   & 0.17      & 0.27     & 0.11 & 0.10    & 0.18       & 0.29     & 0.10 &                      & 0.24   & 0.27      & 0.22     & 0.02 & 0.33    & 0.55       & 0.53     & 0.30 &                      & 0.25   & 0.32      & 0.37     & 0.31 & 0.30    & 0.36       & 0.31     & 0.35 \\
BadT                 & 0.39   & 0.74      & 0.78     & 0.32 & 0.36    & 0.75       & 0.80     & 0.27 &                      & 0.42   & 0.81      & 0.83     & 0.32 & 0.47    & 0.80       & 0.70     & 0.34 &                      & 0.40   & 0.78      & 0.76     & 0.33 & 0.40    & 0.66       & 0.69     & 0.40 \\
CoCoL                & 0.46   & 0.94      & 0.94     & 0.45 & 0.45    & 0.95       & 0.95     & 0.45 &                      & 0.42   & 0.99      & 0.97     & 0.43 & 0.43    & 0.97       & 0.97     & 0.44 &                      & 0.42   & 0.92      & 0.91     & 0.44 & 0.44    & 0.94       & 0.94     & 0.46 \\ \cmidrule(lr){11-18} \cmidrule(lr){19-26}
\multicolumn{9}{c}{BERTdot 5\%}                                                                      & \multicolumn{9}{c}{BERTdot 15\%}                                                                     & \multicolumn{9}{c}{BERTdot 25\%}                                                                     \\
Retrain              & 0.73   & 0.98      & 0.98     & 0.54 & 0.69    & 0.97       & 0.98     & 0.55 &                      & 0.72   & 0.97      & 0.94     & 0.54 & 0.70    & 0.94       & 0.94     & 0.53 &                      & 0.65   & 0.97      & 0.95     & 0.51 & 0.67    & 0.98       & 0.97     & 0.53 \\
CF                   & 0.70   & 0.99      & 0.99     & 0.56 & 0.67    & 0.99       & 0.99     & 0.56 &                      & 0.71   & 0.96      & 0.96     & 0.54 & 0.71    & 0.96       & 0.94     & 0.53 &                      & 0.63   & 0.97      & 0.99     & 0.53 & 0.64    & 0.94       & 0.97     & 0.52 \\
Amnesiac             & 0.02   & 0.01      & 0.02     & 0.01 & 0.01    & 0.01       & 0.02     & 0.01 &                      & 0.06   & 0.02      & 0.07     & 0.01 & 0.08    & 0.07       & 0.10     & 0.12 &                      & 0.01   & 0.08      & 0.09     & 0.07 & 0.04    & 0.10       & 0.06     & 0.07 \\
NegGrad              & 0.04   & 0.72      & 0.58     & 0.30 & 0.10    & 0.63       & 0.66     & 0.15 &                      & 0.04   & 0.70      & 0.67     & 0.31 & 0.07    & 0.69       & 0.69     & 0.11 &                      & 0.34   & 0.70      & 0.60     & 0.20 & 0.10    & 0.59       & 0.63     & 0.24 \\
SSD                  & 0.33   & 0.45      & 0.48     & 0.23 & 0.40    & 0.39       & 0.40     & 0.20 &                      & 0.41   & 0.54      & 0.63     & 0.22 & 0.42    & 0.50       & 0.52     & 0.24 &                      & 0.36   & 0.46      & 0.43     & 0.27 & 0.41    & 0.46       & 0.47     & 0.27 \\
BadT                 & 0.31   & 0.65      & 0.62     & 0.39 & 0.34    & 0.52       & 0.72     & 0.29 &                      & 0.34   & 0.71      & 0.68     & 0.32 & 0.34    & 0.69       & 0.69     & 0.30 &                      & 0.41   & 0.59      & 0.53     & 0.31 & 0.40    & 0.61       & 0.58     & 0.33 \\
CoCoL                & 0.52   & 0.99      & 0.99     & 0.53 & 0.54    & 0.95       & 0.95     & 0.54 &                      & 0.53   & 0.96      & 0.95     & 0.52 & 0.52    & 0.96       & 0.94     & 0.53 &                      & 0.50   & 0.97      & 0.95     & 0.51 & 0.51    & 0.96       & 0.94     & 0.51 \\ \cmidrule(lr){11-18} \cmidrule(lr){19-26}
\multicolumn{9}{c}{ColBERT 5\%}                                                                      & \multicolumn{9}{c}{ColBERT 15\%}                                                                     & \multicolumn{9}{c}{ColBERT 25\%}                                                                     \\
Retrain              & 0.65   & 0.96      & 0.97     & 0.55 & 0.61    & 0.95       & 0.95     & 0.53 &                      & 0.65   & 0.96      & 0.97     & 0.53 & 0.69    & 0.95       & 0.97     & 0.53 &                      & 0.64   & 0.93      & 0.92     & 0.53 & 0.66    & 0.94       & 0.95     & 0.53 \\
CF                   & 0.45   & 1.00      & 0.99     & 0.42 & 0.52    & 0.99       & 0.98     & 0.52 &                      & 0.64   & 1.00      & 0.99     & 0.51 & 0.69    & 0.95       & 0.98     & 0.54 &                      & 0.67   & 0.96      & 0.94     & 0.54 & 0.67    & 0.97       & 0.95     & 0.52 \\
Amnesiac             & 0.03   & 0.04      & 0.10     & 0.03 & 0.13    & 0.25       & 0.13     & 0.09 &                      & 0.02   & 0.04      & 0.06     & 0.10 & 0.07    & 0.04       & 0.10     & 0.09 &                      & 0.05   & 0.07      & 0.01     & 0.12 & 0.07    & 0.03       & 0.02     & 0.14 \\
NegGrad              & 0.25   & 0.50      & 0.58     & 0.32 & 0.35    & 0.37       & 0.41     & 0.33 &                      & 0.32   & 0.33      & 0.35     & 0.25 & 0.35    & 0.24       & 0.26     & 0.30 &                      & 0.30   & 0.35      & 0.39     & 0.33 & 0.31    & 0.30       & 0.33     & 0.35 \\
SSD                  & 0.30   & 0.52      & 0.42     & 0.33 & 0.26    & 0.34       & 0.41     & 0.27 &                      & 0.25   & 0.33      & 0.35     & 0.30 & 0.25    & 0.25       & 0.31     & 0.33 &                      & 0.27   & 0.42      & 0.40     & 0.30 & 0.30    & 0.43       & 0.42     & 0.30 \\
BadT                 & 0.13   & 0.29      & 0.34     & 0.23 & 0.25    & 0.45       & 0.36     & 0.33 &                      & 0.20   & 0.33      & 0.34     & 0.32 & 0.26    & 0.19       & 0.21     & 0.21 &                      & 0.21   & 0.37      & 0.41     & 0.22 & 0.26    & 0.33       & 0.40     & 0.23 \\
CoCoL                & 0.55   & 0.95      & 0.97     & 0.56 & 0.57    & 0.94       & 0.95     & 0.53 &                      & 0.52   & 0.95      & 0.95     & 0.52 & 0.52    & 0.97       & 0.97     & 0.53 &                      & 0.52   & 0.97      & 0.92     & 0.53 & 0.52    & 0.93       & 0.95     & 0.52 \\ \cmidrule(lr){11-18} \cmidrule(lr){19-26}
\multicolumn{9}{c}{PARADE 5\%}                                                                       & \multicolumn{9}{c}{PARADE 15\%}                                                                      & \multicolumn{9}{c}{PARADE 25\%}                                                                      \\
Retrain              & 0.54   & 0.99      & 0.98     & 0.49 & 0.62    & 0.95       & 0.97     & 0.50 &                      & 0.60   & 0.95      & 0.97     & 0.46 & 0.53    & 0.94       & 0.95     & 0.46 &                      & 0.57   & 0.97      & 0.97     & 0.44 & 0.57    & 0.96       & 0.95     & 0.48 \\
CF                   & 0.53   & 1.00      & 0.99     & 0.52 & 0.59    & 0.98       & 0.97     & 0.55 &                      & 0.59   & 0.94      & 0.98     & 0.47 & 0.50    & 0.96       & 0.96     & 0.47 &                      & 0.59   & 0.95      & 0.96     & 0.45 & 0.55    & 0.99       & 0.99     & 0.45 \\
Amnesiac             & 0.01   & 0.00      & 0.01     & 0.00 & 0.02    & 0.02       & 0.03     & 0.07 &                      & 0.03   & 0.01      & 0.10     & 0.10 & 0.03    & 0.03       & 0.06     & 0.01 &                      & 0.07   & 0.07      & 0.10     & 0.01 & 0.07    & 0.09       & 0.09     & 0.05 \\
NegGrad              & 0.31   & 0.62      & 0.54     & 0.20 & 0.20    & 0.31       & 0.42     & 0.34 &                      & 0.35   & 0.60      & 0.61     & 0.37 & 0.32    & 0.63       & 0.62     & 0.38 &                      & 0.33   & 0.61      & 0.58     & 0.35 & 0.33    & 0.60       & 0.52     & 0.31 \\
SSD                  & 0.33   & 0.71      & 0.52     & 0.25 & 0.28    & 0.60       & 0.63     & 0.20 &                      & 0.27   & 0.63      & 0.64     & 0.24 & 0.25    & 0.60       & 0.61     & 0.23 &                      & 0.29   & 0.63      & 0.62     & 0.32 & 0.30    & 0.61       & 0.65     & 0.33 \\
BadT                 & 0.09   & 0.18      & 0.25     & 0.09 & 0.11    & 0.33       & 0.27     & 0.11 &                      & 0.10   & 0.20      & 0.16     & 0.15 & 0.11    & 0.21       & 0.22     & 0.13 &                      & 0.11   & 0.24      & 0.22     & 0.15 & 0.21    & 0.34       & 0.33     & 0.18 \\
CoCoL                & 0.47   & 0.92      & 0.95     & 0.47 & 0.49    & 0.93       & 0.95     & 0.48 &                      & 0.44   & 0.92      & 0.95     & 0.45 & 0.45    & 0.94       & 0.95     & 0.47 &                      & 0.42   & 0.93      & 0.93     & 0.43 & 0.48    & 0.93       & 0.94     & 0.43 \\ \bottomrule

\end{tabular}
}
\label{tab:msmarco_results}
\end{sidewaystable}

\begin{sidewaystable}[htbp]
\caption{
\textbf{Detailed Performance of Unlearning Methods on TREC Dataset.} 
This table mirrors Table~\ref{tab:msmarco_results}  evaluation structure for consistency.}
\resizebox{\textheight}{!}{%
\begin{tabular}{@{}|l|llllllll|l|llllllll|l|llllllll|@{}}
\toprule
\multicolumn{1}{c}{} & \multicolumn{4}{c}{\textbf{Query removal}}  & \multicolumn{4}{c}{Document removal} & \multicolumn{1}{c}{} & \multicolumn{4}{c}{\textbf{Query   removal}}  & \multicolumn{4}{c}{\textbf{Document removal}} & \multicolumn{1}{c}{} & \multicolumn{4}{c}{\textbf{Query removal}}  & \multicolumn{4}{c}{\textbf{Document removal}} \\
                     & F & E & D & T & F  & E  & D & T &                      & F & E & D & T & F  & E  & D & T &                      & F & E & D & T & F & E  & D & T \\ \cmidrule(lr){2-9} \cmidrule(lr){11-18} \cmidrule(lr){19-26}
\multicolumn{9}{c}{BERTcat 5\%}                                                                      & \multicolumn{9}{c}{BERTcat 15\%}                                                                     & \multicolumn{9}{c}{BERTcat 25\%}                                                                     \\
Retrain              & 0.83   & 1.00      & 1.00     & 0.42 & 0.74    & 1.00       & 0.99     & 0.43 &                      & 0.88   & 1.00      & 0.97     & 0.42 & 0.68    & 1.00       & 1.00     & 0.47 &                      & 0.86   & 0.99      & 0.97     & 0.40 & 0.82    & 1.00       & 0.96     & 0.40 \\
CF                   & 0.98   & 1.00      & 1.00     & 0.46 & 0.87    & 1.00       & 1.00     & 0.44 &                      & 0.90   & 1.00      & 0.91     & 0.43 & 0.82    & 0.95       & 1.00     & 0.40 &                      & 0.97   & 0.97      & 0.91     & 0.41 & 0.93    & 1.00       & 1.00     & 0.44 \\
Amnesiac             & 0.02   & 0.01      & 0.01     & 0.01 & 0.01    & 0.01       & 0.02     & 0.01 &                      & 0.02   & 0.01      & 0.02     & 0.01 & 0.01    & 0.01       & 0.02     & 0.01 &                      & 0.02   & 0.01      & 0.02     & 0.01 & 0.01    & 0.02       & 0.02     & 0.01 \\
NegGrad              & 0.03   & 0.84      & 0.63     & 0.32 & 0.09    & 0.50       & 0.58     & 0.17 &                      & 0.04   & 0.83      & 0.63     & 0.30 & 0.09    & 0.63       & 0.53     & 0.15 &                      & 0.03   & 0.70      & 0.55     & 0.34 & 0.08    & 0.60       & 0.73     & 0.15 \\
SSD                  & 0.37   & 0.59      & 0.63     & 0.26 & 0.30    & 0.63       & 0.66     & 0.18 &                      & 0.32   & 0.67      & 0.49     & 0.27 & 0.30    & 0.52       & 0.59     & 0.21 &                      & 0.30   & 0.72      & 0.57     & 0.23 & 0.31    & 0.56       & 0.61     & 0.21 \\
BadT                 & 0.08   & 0.21      & 0.22     & 0.11 & 0.12    & 0.32       & 0.30     & 0.10 &                      & 0.10   & 0.15      & 0.27     & 0.08 & 0.10    & 0.38       & 0.26     & 0.12 &                      & 0.10   & 0.19      & 0.23     & 0.10 & 0.12    & 0.30       & 0.26     & 0.13 \\
CoCoL                & 0.41   & 0.93      & 0.95     & 0.42 & 0.43    & 0.97       & 0.96     & 0.44 &                      & 0.42   & 0.96      & 0.94     & 0.42 & 0.47    & 0.95       & 0.96     & 0.46 &                      & 0.40   & 0.93      & 0.94     & 0.41 & 0.40    & 0.96       & 0.93     & 0.39 \\ \cmidrule(lr){2-9} \cmidrule(lr){11-18} \cmidrule(lr){19-26}
\multicolumn{9}{c}{BERTdot 5\%}                                                                      & \multicolumn{9}{c}{BERTdot 15\%}                                                                     & \multicolumn{9}{c}{BERTdot 25\%}                                                                     \\
Retrain              & 0.63   & 0.99      & 1.00     & 0.43 & 0.65    & 0.98       & 0.99     & 0.45 &                      & 0.69   & 0.99      & 0.99     & 0.47 & 0.72    & 0.99       & 0.99     & 0.46 &                      & 0.68   & 0.98      & 0.98     & 0.45 & 0.70    & 0.99       & 0.99     & 0.43 \\
CF                   & 0.62   & 1.00      & 1.00     & 0.47 & 0.67    & 0.99       & 1.00     & 0.46 &                      & 0.65   & 0.99      & 1.00     & 0.48 & 0.69    & 1.00       & 0.99     & 0.46 &                      & 0.66   & 0.99      & 0.99     & 0.46 & 0.64    & 1.00       & 1.00     & 0.43 \\
Amnesiac             & 0.02   & 0.01      & 0.02     & 0.01 & 0.01    & 0.01       & 0.02     & 0.01 &                      & 0.03   & 0.01      & 0.02     & 0.01 & 0.01    & 0.01       & 0.03     & 0.01 &                      & 0.02   & 0.01      & 0.02     & 0.01 & 0.01    & 0.02       & 0.02     & 0.01 \\
NegGrad              & 0.04   & 0.61      & 0.60     & 0.28 & 0.10    & 0.51       & 0.72     & 0.17 &                      & 0.05   & 0.77      & 0.67     & 0.32 & 0.12    & 0.60       & 0.59     & 0.18 &                      & 0.04   & 0.74      & 0.57     & 0.25 & 0.12    & 0.66       & 0.64     & 0.16 \\
SSD                  & 0.36   & 0.65      & 0.48     & 0.24 & 0.26    & 0.61       & 0.53     & 0.19 &                      & 0.31   & 0.76      & 0.55     & 0.27 & 0.30    & 0.49       & 0.64     & 0.17 &                      & 0.37   & 0.67      & 0.50     & 0.26 & 0.28    & 0.60       & 0.60     & 0.20 \\
BadT                 & 0.09   & 0.19      & 0.28     & 0.10 & 0.10    & 0.32       & 0.29     & 0.13 &                      & 0.08   & 0.17      & 0.25     & 0.08 & 0.12    & 0.34       & 0.26     & 0.10 &                      & 0.10   & 0.19      & 0.30     & 0.09 & 0.13    & 0.40       & 0.33     & 0.11 \\
CoCoL                & 0.43   & 0.95      & 0.97     & 0.43 & 0.45    & 0.95       & 0.96     & 0.44 &                      & 0.46   & 0.98      & 0.97     & 0.46 & 0.46    & 0.97       & 0.96     & 0.46 &                      & 0.45   & 0.96      & 0.97     & 0.45 & 0.43    & 0.97       & 0.95     & 0.42 \\ \cmidrule(lr){2-9} \cmidrule(lr){11-18} \cmidrule(lr){19-26}
\multicolumn{9}{c}{ColBERT 5\%}                                                                      & \multicolumn{9}{c}{ColBERT 15\%}                                                                     & \multicolumn{9}{c}{ColBERT 25\%}                                                                     \\
Retrain              & 0.82   & 0.96      & 0.97     & 0.52 & 0.75    & 0.96       & 0.96     & 0.54 &                      & 0.75   & 0.96      & 0.97     & 0.55 & 0.71    & 0.97       & 0.95     & 0.52 &                      & 0.71   & 0.95      & 0.94     & 0.51 & 0.69    & 0.95       & 0.96     & 0.51 \\
CF                   & 0.79   & 0.97      & 0.99     & 0.53 & 0.72    & 0.96       & 0.99     & 0.55 &                      & 0.72   & 0.97      & 0.96     & 0.57 & 0.70    & 0.96       & 0.96     & 0.51 &                      & 0.68   & 0.96      & 0.96     & 0.50 & 0.67    & 0.96       & 0.97     & 0.52 \\
Amnesiac             & 0.02   & 0.01      & 0.02     & 0.01 & 0.01    & 0.02       & 0.02     & 0.01 &                      & 0.02   & 0.01      & 0.02     & 0.01 & 0.01    & 0.02       & 0.02     & 0.01 &                      & 0.02   & 0.01      & 0.02     & 0.02 & 0.01    & 0.02       & 0.02     & 0.01 \\
NegGrad              & 0.04   & 0.85      & 0.65     & 0.33 & 0.11    & 0.58       & 0.71     & 0.18 &                      & 0.03   & 0.82      & 0.51     & 0.29 & 0.09    & 0.70       & 0.60     & 0.18 &                      & 0.03   & 0.76      & 0.48     & 0.33 & 0.10    & 0.51       & 0.68     & 0.14 \\
SSD                  & 0.29   & 0.59      & 0.58     & 0.22 & 0.31    & 0.63       & 0.68     & 0.20 &                      & 0.40   & 0.61      & 0.44     & 0.29 & 0.34    & 0.72       & 0.59     & 0.17 &                      & 0.35   & 0.68      & 0.54     & 0.28 & 0.30    & 0.57       & 0.55     & 0.19 \\
BadT                 & 0.08   & 0.17      & 0.24     & 0.11 & 0.11    & 0.36       & 0.28     & 0.13 &                      & 0.09   & 0.20      & 0.26     & 0.11 & 0.12    & 0.35       & 0.32     & 0.10 &                      & 0.09   & 0.18      & 0.30     & 0.09 & 0.13    & 0.29       & 0.30     & 0.12 \\
CoCoL                & 0.52   & 0.95      & 0.95     & 0.51 & 0.53    & 0.94       & 0.95     & 0.52 &                      & 0.54   & 0.94      & 0.95     & 0.53 & 0.51    & 0.96       & 0.95     & 0.51 &                      & 0.50   & 0.90      & 0.91     & 0.45 & 0.51    & 0.91       & 0.91     & 0.48 \\ \cmidrule(lr){2-9} \cmidrule(lr){11-18} \cmidrule(lr){19-26}

\multicolumn{9}{c}{PARADE 5\%}                                                                       & \multicolumn{9}{c}{PARADE 15\%}                                                                      & \multicolumn{9}{c}{PARADE 25\%}                                                                      \\
Retrain              & 0.61   & 0.99      & 0.99     & 0.49 & 0.59    & 0.95       & 0.97     & 0.50 &                      & 0.63   & 0.99      & 0.98     & 0.49 & 0.63    & 0.99       & 0.97     & 0.49 &                      & 0.63   & 0.97      & 0.96     & 0.45 & 0.69    & 0.97       & 0.97     & 0.47 \\
CF                   & 0.59   & 1.00      & 1.00     & 0.50 & 0.56    & 0.97       & 0.97     & 0.49 &                      & 0.61   & 1.00      & 1.00     & 0.50 & 0.60    & 0.99       & 0.98     & 0.49 &                      & 0.64   & 0.98      & 0.97     & 0.45 & 0.60    & 0.97       & 0.98     & 0.49 \\
Amnesiac             & 0.02   & 0.01      & 0.02     & 0.01 & 0.01    & 0.02       & 0.02     & 0.01 &                      & 0.02   & 0.01      & 0.02     & 0.02 & 0.01    & 0.02       & 0.02     & 0.01 &                      & 0.02   & 0.01      & 0.02     & 0.01 & 0.01    & 0.01       & 0.02     & 0.01 \\
NegGrad              & 0.04   & 0.71      & 0.65     & 0.31 & 0.11    & 0.74       & 0.55     & 0.15 &                      & 0.04   & 0.78      & 0.60     & 0.29 & 0.12    & 0.75       & 0.63     & 0.17 &                      & 0.04   & 0.70      & 0.56     & 0.31 & 0.10    & 0.60       & 0.70     & 0.15 \\
SSD                  & 0.33   & 0.72      & 0.46     & 0.26 & 0.29    & 0.62       & 0.60     & 0.20 &                      & 0.30   & 0.65      & 0.57     & 0.24 & 0.32    & 0.58       & 0.60     & 0.22 &                      & 0.32   & 0.60      & 0.51     & 0.27 & 0.28    & 0.49       & 0.64     & 0.18 \\
BadT                 & 0.10   & 0.16      & 0.28     & 0.11 & 0.10    & 0.37       & 0.32     & 0.12 &                      & 0.08   & 0.15      & 0.25     & 0.09 & 0.12    & 0.31       & 0.28     & 0.11 &                      & 0.09   & 0.18      & 0.24     & 0.10 & 0.10    & 0.33       & 0.30     & 0.12 \\
CoCoL                & 0.49   & 0.92      & 0.92     & 0.45 & 0.50    & 0.90       & 0.92     & 0.45 &                      & 0.47   & 0.94      & 0.95     & 0.43 & 0.48    & 0.92       & 0.93     & 0.44 &                      & 0.43   & 0.90      & 0.90     & 0.39 & 0.47    & 0.89       & 0.91     & 0.41  \\ \bottomrule
\end{tabular}
}
\label{tab:trec_results}
\end{sidewaystable}

\bibliographystyle{unsrtnat}  
\bibliography{references}

\end{document}